\documentclass[twocolumn,aps,prl,showpacs,raggedbottom,nobalancelastpage,amssymb,superscriptaddress]{revtex4-1}

\usepackage{graphicx}
\usepackage{amsmath}
\usepackage{amssymb}
\usepackage{bm}
\usepackage[usenames]{color}
\usepackage{amsfonts}
\usepackage{appendix}
\usepackage{dsfont}

\def \Fst{1^\textrm{st}}
\def \Snd{2^\textrm{nd}}

\definecolor{Nathanblue}{rgb}{0.,0.24,0.51}

\def\be{\begin{equation}}
\def\ee{\end{equation}}

\def\bs#1{\boldsymbol{#1}}

\begin{document}
\title{Four-Dimensional Quantum Hall Effect with Ultracold Atoms}

\author{H. M. Price}
\affiliation{INO-CNR BEC Center and Dipartimento di Fisica, Universit\`{a} di Trento, I-38123 Povo, Italy}
\author{O. Zilberberg}
\affiliation{Institute for Theoretical Physics, ETH Zurich, 8093 Z{\"u}rich, Switzerland}
\author{T. Ozawa}
\affiliation{INO-CNR BEC Center and Dipartimento di Fisica, Universit\`{a} di Trento, I-38123 Povo, Italy}
\author{I. Carusotto}
\affiliation{INO-CNR BEC Center and Dipartimento di Fisica, Universit\`{a} di Trento, I-38123 Povo, Italy}
\author{N. Goldman}
\email[]{ngoldman@ulb.ac.be}
\affiliation{CENOLI, Facult{\'e} des Sciences, Universit{\'e} Libre de Bruxelles (U.L.B.), B-1050 Brussels, Belgium}


\begin{abstract}
We propose a realistic scheme to detect the 4D quantum Hall effect using ultracold atoms. Based on contemporary technology, motion along a synthetic fourth dimension can be accomplished through controlled transitions between internal states of atoms arranged in a 3D optical lattice. From a semi-classical analysis, we identify the linear and non-linear quantized current responses of our 4D model, relating these to the topology of the Bloch bands. We then propose experimental protocols, based on current or center-of-mass-drift measurements, to extract the topological $\Snd$ Chern number. Our proposal sets the stage for the exploration of novel topological phases in higher dimensions.
\end{abstract}


\maketitle

In the past decade, there have been rapid developments in the control of coherent quantum systems. These advances have led to the experimental exploration of fundamental quantum mechanical concepts, many of which were previously only of theoretical interest. Specifically, fascinating artificial systems can now be engineered with atoms in optical lattices, where topological order, gauge structures, disorder, and interactions can all be tuned and probed in novel ways~\cite{Bloch:2008gl,Dalibard2011,Goldman:2014bv}. One such important recent achievement has been the realization of artificial magnetic fields and topological Bloch bands using time-modulated two-dimensional (2D) optical lattices~\cite{Struck, aidelsburger2013,miyake2013,jotzu2014,Aidelsburger:2015}. 

Optical lattices can be tailored to realize lattice geometries of varying spatial dimensions $d_{\text{real}}\!=\!1,2,3$. Moreover, increasing control over atomic internal states~\cite{Jaksch2005,Bloch:2008gl,Weitenberg:2011,Dalibard2011,Goldman:2014bv} has extended design flexibility as ``motion" along an auxiliary (synthetic) dimension can be mimicked by laser-induced transitions between internal states \cite{Boada2012,Celi:2014, Cooper:2015, Barbarino:2015}. This effectively generates dynamics within synthetic geometries with $d\!=\!(d_{\text{real}}\!+\!1)$ ``spatial" dimensions, in addition to the usual time dimension. This was recently demonstrated in experiments, where 2D physics was explored with atoms in a one-dimensional optical lattice with a synthetic dimension \cite{Mancini:2015,Stuhl:2015}. These developments naturally open up the possibility of emulating systems with higher dimensions $d\!>\!3$. For example, adding a synthetic dimension to a three-dimensional (3D) optical lattice would simulate a system in four dimensions (4D). Importantly, artificial gauge potentials~\cite{Dalibard2011,Goldman:2014bv} can naturally be introduced in atomic systems with synthetic dimensions~\cite{Celi:2014} in the form of Peierls phase-factors. These gauge structures can be finely tuned through the laser-coupling, allowing experiments to explore topologically-nontrivial energy bands in higher spatial dimensions.

 In the fast-expanding field of topological phases of matter, the energy bands of a system are associated with topological indices that are robust to continuous deformations. Nontrivial indices are usually associated with interesting boundary phenomena, quantized responses, and exotic quasiparticles \cite{RMP_TI,RMP_TI2}. In the 2D quantum Hall (QH) effect, the quantized Hall conductance is related to the sum of the $\Fst$ Chern numbers (1CNs) of the filled energy bands~\cite{TKNN}. In 4D, even more intriguing quantum Hall phases may exist, as first predicted for time-reversal (TR) symmetric systems \cite{IQHE4D,QiZhang}. In the 4D QH effect, energy bands possess an additional nontrivial topological index, the $\Snd$ Chern number (2CN), leading to a nonlinear quantized response~\cite{Yang78,IQHE4D,QiZhang}. Both the 2D and 4D QH effects exhibit a variety of exotic strongly-correlated phases in the presence of interactions~\cite{IQHE4D,FQHE_review}. Hence, 4D models with non-vanishing 2CNs have attracted much theoretical attention~\cite{QiZhang,LiZhang2012,Li:2012, Li:2013, Edge:2012,Kraus:2013}.

In this Letter, we describe a concrete proposal for realizing the 4D QH effect using ultracold gases in optical lattices. We present a semi-classical analysis predicting the general transport equations in 4D setups: this includes a non-linear response, related to the 2CN, but also a linear response, associated with an exotic 2D QH effect for the current flowing across 2D planes within the 4D system. We then propose realistic protocols through which these responses could be measured experimentally. 

{\it Semi-classical Analysis:} We consider particles of charge $q\!=\!-1$~\cite{footnote_field}, moving in a lattice system and prepared in a non-degenerate energy band $\mathcal{E}(\bs k)$. The geometrical properties of the Bloch band's eigenstates $| u \rangle$ are encoded in the Berry curvature 2-form~\cite{Xiao2010}, $\Omega\!=\!(1/2) \Omega^{\mu \nu}(\bs k) \text{d}k_{\mu}\! \wedge \! \text{d}k_{\nu}$, with components $\Omega^{\mu \nu}\!=\! i \left (\langle \partial_{k_\mu } u \vert \partial_{k_\nu } u \rangle \!-\! \langle \partial_{k_\nu } u \vert \partial_{k_\mu } u \rangle \right )$.  The indices $\mu$ run over spatial coordinates with Einstein summation convention, and we set $\hbar\!=\!1$ unless otherwise stated. Specifically, we are interested in transport under external (weak) electric $\bs E= E_{\mu} \bs{e}^{\mu}$ and magnetic fields, where $B_{\mu \nu}\!=\!\partial_{\mu}A_{\nu}\!-\!\partial_{\nu}A_{\mu}$ is the magnetic field strength and $\bs A\!=\! A_{\mu} \bs{e}^{\mu}$ is the vector potential. 
We assume that any strong magnetic fields are directly incorporated into the Bloch band via the (magnetic) eigenstates $| u \rangle$~\cite{chang2}. 
Considering for now a system of arbitrary dimensions ($d \ge 2$), we write the semi-classical equations of motion for a wave-packet as~\cite{footnote_semiclassics, chang1, chang2, Xiao2010, Gao}
\begin{align}
&\dot r^{\mu}(\bs k)=\frac{\partial \mathcal{E}(\bs k)}{\partial k_{\mu}} - \dot k_{\nu} \Omega^{\mu \nu}(\bs k), \label{eq:firstline}   \\
&\dot k_{\mu}=- E_{\mu} -\dot r^{\nu} B_{\mu \nu}, \label{eq:secondline}
\end{align}
where $\bs r\!=\! r^{\mu} \bs{e}_{\mu}$ [resp. $\bs k\!=\! k_{\mu} \bs{e}^{\mu}$] denotes the mean position [resp. quasi-momentum] of the wave-packet. We stress that the external fields $E_{\mu}$ and $B_{\mu \nu}$ are sufficiently weak perturbations that, at this level, the wave-packet adiabatically follows the band $\mathcal{E}(\bs k)$~\cite{Sup_Mat}. Next, we insert Eq.\eqref{eq:secondline} into Eq.\eqref{eq:firstline}, in order to derive a transport equation relating the mean velocity to the external fields. To second order in the perturbations, this yields~\cite{Sup_Mat}
\begin{align}
\dot r^{\mu}(\bs k)&=\frac{\partial \mathcal{E}(\bs k)}{\partial k_{\mu}} + E_{\nu} \Omega^{\mu \nu}(\bs k)
+ E_{\delta}B_{\nu \gamma} \Omega^{\gamma \delta}(\bs k) \Omega^{\mu \nu}(\bs k) \notag \\
&+ \left( \frac{\partial \mathcal{E}(\bs k)}{\partial k_{\gamma}}  +  \frac{\partial {\mathcal{E}}(\bs k)}{\partial k_\alpha }   \Omega^{\gamma \delta} (\bs k) B_{ \delta \alpha}  \right)   \Omega^{\mu \nu} (\bs k) B_{\nu \gamma} .
 \label{transport_general}
\end{align}
In addition to the group velocity of the band $\partial_{k_{\mu}}  \mathcal{E}$, the first line of \eqref{transport_general} includes two important response terms. The first of these corresponds to a transverse drift induced by the electric field, which is responsible for the QH effect in 2D; the second reveals a non-linear (electromagnetic) response, which is quadratic in the Berry curvature. In this Letter, we are first interested in transport of fermions completely filling the band $ \mathcal{E}(\bs k)$. In this case, neither the group velocity nor the terms in the second line of \eqref{transport_general} contribute to the response \cite{Sup_Mat}. Anticipating that the non-linear term in the first line of \eqref{transport_general} will be crucial when $d \!\ge\! 4$, as suggested by topological field theory~\cite{IQHE4D,QiZhang}, we now explicitly consider $d\!=\!4$.

{\it 4D Transport Equations and Topological Invariants}: In the semi-classical approach, special care is required when evaluating the contribution of all the Bloch states in a given Bloch band \cite{Xiao2010}. An integration performed over phase-space involves the usual density of states $(1/2\pi)^d$ whenever either the external magnetic field or the Berry curvature is absent. However, in the presence of both quantities, the physical position and momentum must  both be distinguished from the canonical coordinates, leading to a corresponding modification of the density of states \cite{Xiao_ED, Duval, Bliokh}. In 4D, we find that the sum over all Bloch states can be accurately replaced by the integral~\cite{Sup_Mat}
\begin{align}
\sum_{\bs k} \equiv \frac{V}{(2 \pi)^4} &\int_{\mathbb{T}^4} \Biggl [1 + \frac{1}{2} B_{\mu \nu} \Omega^{\mu \nu} \label{canonical_transf} \\
&+\frac{1}{64} \left (\varepsilon^{\alpha \beta \gamma \delta} B_{\alpha \beta} B_{\gamma \delta}   \right )\left (\varepsilon_{\mu \nu \lambda \rho} \Omega^{\mu \nu} \Omega^{\lambda \rho}   \right )   \Biggr ]  \text{d}^4 k  , \notag
\end{align}
where indices now take values $\mu\!=\!x,y,z,w$, $V$ is the system's volume, $\varepsilon^{\alpha \beta \gamma \delta}$ is the 4D Levi-Civita symbol, and where the integration is over the first (magnetic) Brillouin zone ($\mathbb{T}^4$). One readily verifies that Eq.~\eqref{canonical_transf} generalizes the  phase-space integration previously derived for 3D systems \cite{Xiao_ED}. Combining Eq. \eqref{canonical_transf} with Eq. \eqref{transport_general} one obtains the transport for a filled band, i.e. defining the current density  $j^{\mu} = \sum_{\bs k} \dot r^{\mu}(\bs k)/V$ yields~\cite{Sup_Mat}
\begin{align}
&j^{\mu}=E_{\nu} \frac{1}{(2\pi)^4} \int_{\mathbb{T}^4} \Omega^{\mu \nu} \text{d}^4k + \frac{\nu_2}{4 \pi^2} \varepsilon^{\mu \alpha \beta \nu} E_{\nu} B_{\alpha \beta} , \label{central_transport_equation}
\end{align}
introducing the 2CN \cite{Avron1988,IQHE4D,QiZhang, footnote_chern1, Fukui:2008}
\begin{align}
\nu_2\!&=\!\frac{1}{8 \pi^2} \int_{\mathbb{T}^4}  \Omega \wedge \Omega \, \, \in \mathbb{Z}, \notag \\
&=\!\frac{1}{4 \pi^2} \int_{\mathbb{T}^4}  \Omega^{xy}\Omega^{zw}\!+\! \Omega^{wx}\Omega^{zy}\!+\! \Omega^{zx}\Omega^{yw} \text{d}^4k.\label{second_chern}
\end{align}
The transport equation (\ref{central_transport_equation}) is a central result of this work; it was derived from the semi-classical equations of motion \eqref{eq:firstline}-\eqref{eq:secondline}, without any assumptions about the system beyond the non-degeneracy of the Bloch band. The only two contributions for the filled band are a linear response, reminiscent of the 2D quantum Hall effect~\cite{TKNN}, and a non-linear response proportional to the 2CN. We emphasise that the latter effect only takes place in systems of dimension $d \!\ge\! 4$, as formally imposed by the Levi-Civita symbol. This 4D QH effect was originally derived from a topological-field-theory approach \cite{IQHE4D,QiZhang} for TR-symmetric spinful systems, where the linear response vanishes by symmetry~\cite{footnote_symmetry}. We note that 2CNs were previously studied from semi-classics in the polarization of inhomogenous crystals~\cite{Xiao:2009}. The transport equation \eqref{central_transport_equation} can also be derived from a Streda-formula approach~\cite{Sup_Mat}. 

\begin{figure}[t!]
\includegraphics[width=9cm]{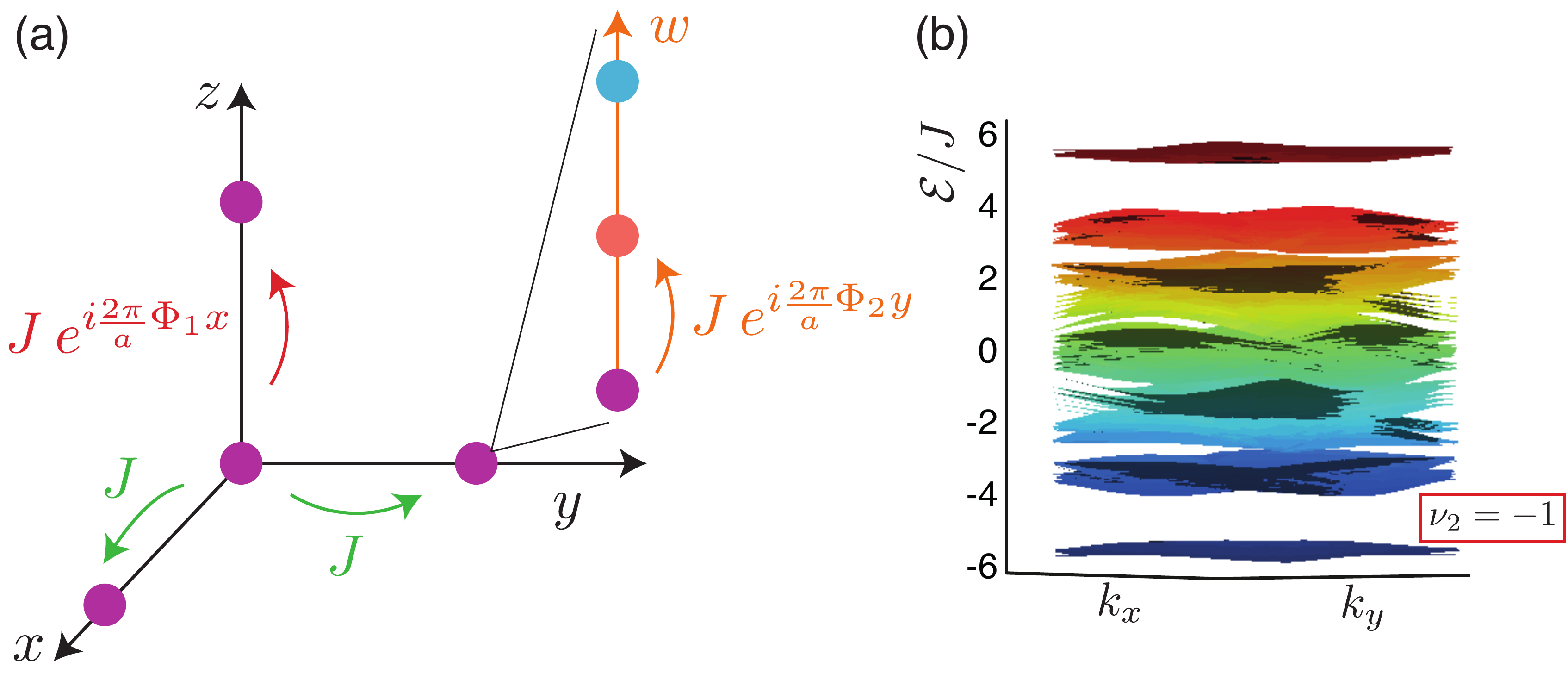}
\vspace{-0.cm} \caption{(a) The 4D lattice in the absence of perturbing fields ($E_{\mu},B_{\mu \nu}\!=\!0$). Atoms move in a 3D optical lattice, with hopping amplitude $J$. A Hofstadter model~\cite{Hofstadter} is realized in the $x\!-\!z$ plane, with flux $2 \pi \Phi_1$ per unit cell, from $x$-dependent Peierls phase-factors in the hopping along the $z$ direction~\cite{aidelsburger2013,miyake2013,Aidelsburger:2015}. A fourth (synthetic) dimension \cite{Boada2012,Celi:2014}, with coordinate $w$, is accessible at each lattice site, as Raman coupling induces internal-state transitions \cite{Sup_Mat}. A second Hofstadter model, with flux $2 \pi \Phi_2$, is realized in the $y\!-\!w$ plane by adding a $y$-dependent phase-factor to the internal-state transitions, i.e.~by aligning the Raman wave-vector along the $y$ direction~\cite{Celi:2014,Mancini:2015,Stuhl:2015}. (b) Energy spectrum $\mathcal{E}(k_x,k_y)$ for $\Phi_{1,2}\!=\!1/4$, and for many values  $k_{z,w}$ over the Brillouin zone; the 2CN of the lowest (nondegenerate) band is indicated.}\label{Fig_1}
\end{figure}

{\it Realistic Physical System:} We now describe our experimental proposal for observing 4D QH physics [Eq.~\eqref{central_transport_equation}] based on cold-atom technology.  As is clear from Eq.~\eqref{second_chern}, a non-trivial 2CN requires that the Berry curvature has non-zero components in ``disconnected" planes, e.g. $\Omega^{zx}, \Omega^{yw}\! \ne\! 0$. This sets a constraint on the minimal models displaying non-trivial 4D bands. Inspired by Ref. \cite{Kraus:2013}, we propose to build a 4D lattice in which the $x\!-\!z$ and  $y\!-\!w$ planes are penetrated by uniform magnetic fluxes $\Phi_{1,2}$, respectively. This corresponds to two copies of the Hofstadter model \cite{Hofstadter}  defined in disconnected planes. The Hamiltonian is thus of the tight-binding form
\begin{align}
\hat H=-J &\sum_{\bs r} c^{\dagger}_{\bs r + a\bs e_x}c_{\bs r}+c^{\dagger}_{\bs r + a\bs e_y}c_{\bs r} \label{4D_ham} \\
& + e^{i 2 \pi \Phi_1 x/a} c^{\dagger}_{\bs r + a\bs e_z}c_{\bs r}+ e^{i 2 \pi \Phi_2 y/a} c^{\dagger}_{\bs r + a\bs e_w}c_{\bs r} + \text{h.c.} \notag ,
\end{align}
where $c^{\dagger}_{\bs r}$ creates a fermion at lattice site $\bs r\!=\!(x,y,z,w)$, $a$ is the lattice spacing, $J$ is the hopping amplitude [Fig. \ref{Fig_1}(a)]. This model has two main ingredients: (i) $x$-dependent Peierls phase-factors for tunneling along the $z$ direction, generating a uniform flux $\Phi_1$ in the $x\!-\!z$ plane, and (ii) $y$-dependent Peierls phase-factors for tunneling along the $w$ direction, creating a uniform flux $\Phi_2$  in the $y\!-\!w$ plane. The first ingredient (i) can be engineered using the method implemented in Refs. \cite{aidelsburger2013,miyake2013,Aidelsburger:2015}, i.e., by combining a superlattice (or a Wannier-Stark ladder) along the $z$ direction with a resonant time-modulation of the optical-lattice  potential~\cite{Kolovsky:2011,Bermudez:2011prl,Goldman:2015}. The second ingredient (ii) requires an extra (synthetic) dimension \cite{Boada2012,Celi:2014, Mancini:2015,Stuhl:2015}, which can be simulated by adjusting two-photon Raman coupling to induce transitions between different internal atomic states~\cite{Sup_Mat}. The Raman wave-vector should be aligned along the $y$ direction so as to realize  $y$-dependent Peierls phase-factors [Fig. \ref{Fig_1}(a)], as was successfully implemented in Refs.~\cite{Mancini:2015,Stuhl:2015}. In general, hopping processes along the synthetic dimension are inhomogeneous~\cite{Celi:2014,Sup_Mat}, but we verified that this does not significantly alter the topological band properties studied below \cite{Sup_Mat}. Hence, for the sake of simplicity, we assume the 4D system~\eqref{4D_ham} is isotropic with a single hopping rate $J$ for all hopping processes.

The bulk energy spectrum $\mathcal{E}(\bs k)$ of the model \eqref{4D_ham}
is reminiscent of the two underlying (2D) Hofstadter models defined in the $x\!-\!z$ and $y\!-\!w$ planes  \cite{Kraus:2013,Sup_Mat}. Specifically, the spectrum corresponds to the Minkowski sum $\mathcal{E}\!=\!\{ E_1 +E_2 \vert E_1\in \mathcal{E}_{xz}, E_2 \in \mathcal{E}_{yw} \}$, where $\mathcal{E}_{xz}$ and  $\mathcal{E}_{yw}$ denote the spectra associated with the 2D Hofstadter models \cite{Sup_Mat}. This rich 4D spectrum has a lowest band $\mathcal{E}_1(\bs k)$, which is generally non-degenerate and, for suitable fluxes of the form $\Phi_1\!=\!1/q_1$ and $\Phi_2\!=\!1/q_2$ with $q_1,q_2 \in \mathbb{Z}$, can be well isolated from higher-energy bands. This happens, for example, for fluxes $\Phi_{1,2}\!=\!1/4$ where the lowest band $\mathcal{E}_1(\bs k)$ has a  large flatness ratio $\Delta/W\!\approx 3$, for bandwidth $W$ and lowest bulk gap $\Delta\!\approx\!1.3J$ (Fig.~\ref{Fig_1}(b)). We henceforth focus on this choice of $\Phi_{1,2}$, as it is relevant to recent experiments on the 2D Hofstadter model \cite{aidelsburger2013,Aidelsburger:2015}.

The bulk spectrum also inherits the topological properties of the underlying Hofstadter models~\cite{Kraus:2013,Sup_Mat}. The lowest band $\mathcal{E}_1(\bs k)$ is characterized by a non-zero 2CN~\cite{footnote_chern}, which can be factorized as $\nu_2\!=\!\nu_1^{zx}\! \times \! \nu_1^{yw}$, where we introduced the 1CNs associated with the lowest bands of the 2D spectra $\mathcal{E}_{xz}$ and $\mathcal{E}_{yw}$, respectively~\cite{Sup_Mat}. Note that the 4D Berry curvature $\Omega^{\mu \nu}$ and the 2CN $\nu_2$ of the lowest band  can be evaluated numerically, with a 4D generalization  \cite{Sup_Mat} of the method developed in Ref.~\cite{Fukui:2005}.

{\it Response of the System to External Fields:} We now investigate the transport equations [Eq.~\eqref{central_transport_equation}] with external ``electric" $E_{\mu}$ and ``magnetic"  $B_{\mu \nu}$ fields. As the atoms are neutral, the ``electric" field $E_{\mu}$ corresponds here to a linear gradient, which can either be magnetically \cite{aidelsburger2013,miyake2013} or optically \cite{Aidelsburger:2015} created. The perturbing ``magnetic"  field $B_{\mu \nu}$ can be generated by engineering additional Peierls phase-factors. For the sake of experimental simplicity, we choose the weak field $B_{\mu \nu}$ to have a single non-zero component, $B_{zw}\!= - \!2 \pi \tilde{\Phi}/a^2$, as is readily achieved by adding an extra $z$-dependent phase-factor to internal-state transitions~\cite{footnote_phasefactor}; this occurs in our proposed system if the Raman wave-vector is tilted in the $y\!-\!z$ plane. To isolate the linear and non-linear responses, we align the ``electric" field along the $y$ direction, $\bs E\!=\!E_y \bs e^y$. The transport equations \eqref{central_transport_equation} are then explicitly $j^y=j^z=0$, and
\begin{align}
&j^x=  \frac{\nu_2}{4 \pi^2} E_{y} B_{zw} = -  \frac{\nu_2}{2 \pi a^2} E_{y} \tilde \Phi \,, \label{non_lin_response}\\
&j^{w}=E_{y} \frac{1}{(2\pi)^4} \int_{\mathbb{T}^4} \Omega^{w y} \text{d}^4k = - \frac{\nu_1^{yw} }{2 \pi a^2 q_1} E_{y} . \label{lin_response}
\end{align}
 Importantly, Eq.~\eqref{non_lin_response} reveals a genuine non-linear 4D-QH response along the $x$ direction. In Eq.~\eqref{lin_response}, we have highlighted the  1CN $\nu_1^{yw}$ associated with the $y\!-\!w$ plane~\cite{footnote_1CN}, and have used that the flux is $\Phi_1\!=\!1/q_1$ in the $x\!-\!z$ plane. Hence,  Eq.~\eqref{lin_response} shows that an exotic 2D QH effect takes place in the $y\!-\!w$ plane: defining the Hall  conductivity in this plane as $\sigma_\text{wy}=j^w_{\text{2D}}/E_y=a^2 j^w/E_y$, we find 
\be
\sigma_{\text{wy}}=-(e^2/h) \,  \nu_1^{yw}/q_1, \qquad q_1, \nu_1^{yw} \in \mathbb{Z}, \label{fractional}
\ee
where the conductivity quantum $e^2/h$ is re-introduced. Such a ``fractional QH effect"  is allowed in non-interacting systems provided the associated 2D plane is embedded in a higher-dimensional system \cite{RMP_TI2}, as recently shown in 3D topological insulators \cite{Xu:2014}. Here, the fractionalization of the Hall conductivity $\sigma_{\text{wy}}$ reflects the change in size of the magnetic Brillouin zone due to the flux $\Phi_1\!=\!1/q_1$ penetrating the disconnected plane $x\!-\!z$.

\begin{figure}[t!]
\includegraphics[width=8.8cm]{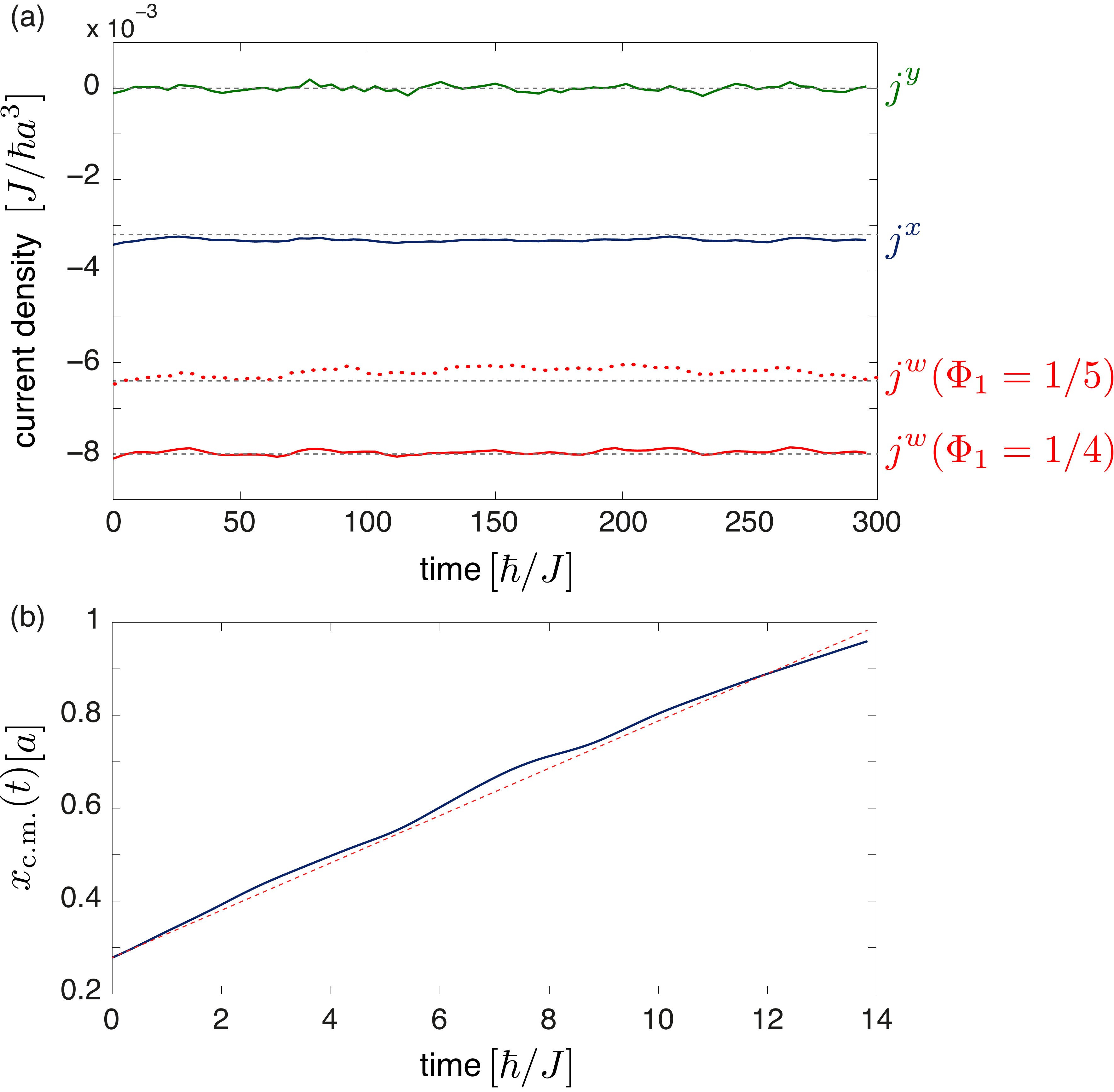}
\vspace{-0.cm} \caption{(a) Current density $j^{\mu} (t)$ after ramping up the ``electric" field to $E_y\!=\!-0.2 J/a$. The fluxes were taken $\Phi_{1,2}\!=\!1/4$, and the Fermi energy $E_{\text{F}}\!=\!-5J$ was set in the first bulk energy gap. The perturbing magnetic field was chosen as $B_{zw}\!=-\!2 \pi \tilde \Phi /a^2\!=-\!\pi/5a^2$. Simulations were performed on a $8 \times 10\times 10\times 4$ lattice, with periodic boundary conditions along the $x,z,w$ directions. The component $j^w$ was also calculated for $\Phi_1\!=\!1/5$ (red dotted line). Black lines show predictions [Eqs.~\eqref{non_lin_response}-\eqref{lin_response}] for $\nu_2\!=\!\nu_1^{yw}\!=-\!1$. (b) Center-of-mass trajectory $x_{\text{c.m.}} (t)$ after ramping up the ``electric" field to $E_y\!=\!0.2 J/a$; the non-linear response is in the opposite direction to (a) as $E_y$ is reversed. Simulations are for a  $12 \times 12\times 10\times 4$ lattice, with periodic boundary conditions along the $z,w$ directions. The initial radius of the cloud was $r_0\!=\!4a$ in the $x\!-\!y$ plane. The fluxes  are $\Phi_{1,2}\!=\!1/4$, with all other parameters the same as in (a). The red dotted curve shows the predicted drift, $v^x_{\text{c.m.}}\!=\! j^x \times 16 a^4\!=\! -8 a^2 \nu_2 E_y \tilde \Phi/\pi$, for $\nu_2\!=\!-1$.}\label{Fig_2}
\end{figure}

{\it Numerical Calculations:} In order to test the predictions in Eqs.~\eqref{non_lin_response}-\eqref{lin_response}, we have numerically calculated the current density $j^{\mu}$ in a 4D  lattice-system filling the lowest bulk band. The plots in Fig. \ref{Fig_2}(a) show the quasi-stationary current density $j^{\mu}$ obtained after ramping up the field $E_y$ to the final value $E_y\!=\!-0.2 J/a$, which was chosen so as to limit inter-band transitions~\cite{Price:2012,Dauphin:2013,Aidelsburger:2015}. Comparing these numerical results to Eqs.~\eqref{non_lin_response}-\eqref{lin_response} allows one to extract the topological indices: we find  $\nu_2 \!\approx \!-1.07$ and $\nu_1^{yw} \!\approx\!-1.03$, in agreement with the topological band structure ($\nu_2\!=\!\nu_1^{yw}\!=-\!1$). These simulations also validate the $\Phi_1$-dependence in Eqs.~\eqref{lin_response}-\eqref{fractional}, as we see that $j^w$ is indeed reduced by a factor $4/5$ when changing the flux $\Phi_1\!=\!1/4 \!\rightarrow\! 1/5$ in the $x\!-\!z$ plane. 

While current measurements could be performed in cold-atom experiments \cite{Bloch_discussion, Brantut}, e.g.~by generalizing the method of Ref.~\cite{Atala:2014}, a more accessible probe of the topological transport would be the observation of the center-of-mass (COM) motion of the cloud~\cite{Dauphin:2013,Aidelsburger:2015}. The COM velocity for this system is $v_{\text{c.m.}}^{\mu}\!=\!j^{\mu} V_{\text{cell}}$, where $j^{\mu}$ is given in Eqs.~\eqref{non_lin_response}-\eqref{lin_response}, and $V_{\text{cell}}$ is the (magnetic) unit cell volume~\cite{footnote_volume, Dauphin:2013,Aidelsburger:2015}, which is $V_{\text{cell}}\!=\!16 a^4$ for $\Phi_{1,2}\!=\!\!1/4$. For a perturbing flux $\tilde \Phi\!=\! 1/10$, we find that the 2CN response should lead to a drift along the $x$ direction with $v_{\text{c.m.}}^{x}\!\approx\! 2 a/T_{\text{B}}$, where $T_{\text{B}}\!=\!2 \pi/a E_y\!\approx\! 50\text{ms}$ is a typical Bloch oscillation time~\cite{Aidelsburger:2015}. Such a cloud displacement is of the same order as the one reported in the recent  1CN-measurement \cite{Aidelsburger:2015}. We analyzed the COM displacement $x_{\text{c.m.}} (t)$ by simulating a 4D system with open boundary conditions in the $x,y$ directions. Numerically, the cloud is initially confined in the presence of all fluxes, then the confinement is removed and the ``electric" field is ramped up~\cite{Dauphin:2013}. From the  COM drift in Fig. \ref{Fig_2}(b), we extract a realistic ``experimental" value for the 2CN: $\nu_2 \!\approx \!-0.98$. 

{\it Discussion:} The COM-response, presented in Fig. \ref{Fig_2}(b) for a Fermi system initially filling the lowest energy band, is strictly identical to that obtained if the band is uniformly, but only partially, filled \cite{Sup_Mat}. This situation typically occurs when a thermal gas is loaded into an energy band with large flatness ratio~\cite{Aidelsburger:2015}, such as the lowest band in Fig.~\ref{Fig_1}(b). Consequently, the 4D QH physics discussed here could equally be explored using Fermi or Bose gases~\cite{Aidelsburger:2015}. We note that other external-field configurations could be chosen to fully test the 4D physics in Eq.~\eqref{central_transport_equation}. We also stress that an accurate measurement of the 2CN requires a large number of internal states~\cite{Sup_Mat}; this can be solved experimentally by imposing periodic boundary conditions along the synthetic dimension, as could be realized with an extra Raman coupling~\cite{Celi:2014}. Finally, we note that the topological gap $\Delta\!\approx\! 1.3J$ is of the same order as in~\cite{Aidelsburger:2015}; this requires low but achievable temperatures to extract the band's topology (see~\cite{Aidelsburger:2015} for methods to extract the Chern number in the presence of band repopulation, and~\cite{heating_ref} on heating sources).

{\it Conclusion:} We have proposed a realistic physical platform to observe the 4D QH effect in cold-atom transport experiments \cite{Price:2012,Dauphin:2013,jotzu2014,Aidelsburger:2015}. We have highlighted the minimal requirements for measuring a non-trivial 2CN in an atomic system, exploiting Raman transitions between internal states~\cite{Celi:2014}. Our proposal sets the stage for the future experimental exploration of higher-dimensional topological phases. The principles of our set-up could also be extended beyond cold atoms, for example, to photonic systems~\cite{4Dphotons:2015}. Realizing such 4D systems experimentally will be especially intriguing as they may harbor exotic collective excitations~\cite{IQHE4D}.

\acknowledgments

We thank N. R. Cooper, J. Dalibard, A. Celi, I. Bloch, M. Aidelsburger, C. Weitenberg, L. Fallani, P. Massignan, M. Lewenstein, A.~Dauphin, P. Gaspard, J. Catani, C.~Sias, L. Mazza, K. Kraus and Z. Ringel for fruitful discussions. %
H.M.P., T.O. and I.C. were funded by ERC through the QGBE grant, by the EU-FET Proactive grant AQuS (Project No. 640800), and by Provincia Autonoma di Trento, partially through the project ``On silicon chip quantum optics for quantum computing and secure communications - SiQuro". O.Z. acknowledges the Swiss National Foundation for financial support. 
N.G. is financed by the FRS-FNRS Belgium and by the BSPO under the PAI project P7/18 DYGEST.%


\begin{thebibliography}{39}
\bibitem{footnote_chern1} This is the 2CN defined from the Chern character rather than the Chern class \cite{Nakahara, Ryu}.
\bibitem{footnote_phasefactor}Following our convention for the Peierls phase-factors [see Fig.~\ref{Fig_1}(a)], the magnetic field strength felt by a particle of charge $q\!=\!-1$ is $B_{\mu \nu}\!=\! - 2 \pi \Phi_{\mu \nu}/a^2$; in the text, the strong fluxes are $\Phi_1\!=\!\Phi_{xz}$, $\Phi_2\!=\!\Phi_{yw}$ and the perturbing flux is $\tilde{\Phi}\!=\! \Phi_{zw}$.
\bibitem{footnote_1CN} Note that $\nu_1^{yw} \!=\! \frac{1}{2\pi} \int \Omega^{yw} \text{d}k_y \text{d}k_w= -\frac{1}{2\pi} \int \Omega^{wy} \text{d}k_y \text{d}k_w$ where the integral is over the $k_y\!-\!k_w$ Brillouin zone~\cite{Sup_Mat}.
\bibitem{footnote_volume} The current density is ${\bs j}={\bs v}_{\text{tot}}/V$, while the COM velocity is ${\bs v}_\text{c.m.}={\bs v}_{\text{tot}}/N_{\text{tot}}$, where $N_{\text{tot}}$ is the particle number and ${\bs v}_{\text{tot}}$ is the total velocity. Combining these, ${\bs v}_\text{c.m.}= {\bs j}/ n$ with $n$ being the particle density of the filled band. For this case, $n=1/ V_{\text{cell}}$, and so ${\bs v}_\text{c.m.}={\bs j} V_{\text{cell}}$.
\end{thebibliography}


\end{document}


\title{Supplemental Material of \\ Four-Dimensional Quantum Hall Effect with Ultracold Atoms}

\author{H. M. Price}
\affiliation{INO-CNR BEC Center and Dipartimento di Fisica, Universit\`{a} di Trento, I-38123 Povo, Italy}
\author{O. Zilberberg}
\affiliation{Institute for Theoretical Physics, ETH Zurich, 8093 Z{\"u}rich, Switzerland}
\author{T. Ozawa}
\affiliation{INO-CNR BEC Center and Dipartimento di Fisica, Universit\`{a} di Trento, I-38123 Povo, Italy}
\author{I. Carusotto}
\affiliation{INO-CNR BEC Center and Dipartimento di Fisica, Universit\`{a} di Trento, I-38123 Povo, Italy}
\author{N. Goldman}
\affiliation{CENOLI, Facult{\'e} des Sciences, Universit{\'e} Libre de Bruxelles (U.L.B.), B-1050 Brussels, Belgium}




\maketitle

\section{Current Density and Transport Equations  from Second-Order semi-classics}
\label{sec:currentdensity}

In this Appendix, we show how the 4D current density (Eq. 5 in the main text) is derived from the semi-classical dynamics of a wave packet. As introduced above, we consider particles of charge $q=-1$ moving in a lattice in the presence of an external electric field $\bs E= E_{\mu} \bs{e}^{\mu}$ and magnetic field strength $B_{\mu \nu}\!=\!\partial_{\mu}A_{\nu}\!-\!\partial_{\nu}A_{\mu}$, where $\bs A= A_{\mu} \bs{e}^{\mu}$ denotes the electromagnetic  vector potential and where indices run over the spatial coordinates (Einstein summation convention). We assume that these external fields are weak perturbations, which are both time-independent and spatially-uniform. We also consider that any strong magnetic fields are incorporated directly into the (magnetic) eigenstates $| {u}_n \rangle $ of the underlying Bloch bands with band index $n$~\cite{chang1, chang2}. We stress that the following discussion is general and also applies to neutral particles, in which case the external fields are artificial~\cite{Dalibard2011, Goldman:2014bv}.  

The wave packet is constructed with a centre of mass at position $\bs r_c\!=\! r_c^{\mu} \bs{e}_{\mu}$ and quasi-momentum $\bs k^c\!=\! k^c_{\mu} \bs{e}^{\mu}$. Provided that the external magnetic vector potential ${\bs A}({\bs r})$ varies on a length scale much larger than the spatial extent of the wave packet, the full quantum Hamiltonian, including the perturbing electromagnetic fields, can be expanded around ${\bs r}_c$ as \cite{chang1, chang2, XiaoPol, Gao, Gao2}
\begin{align}
\hat{H} \approx \hat{H}_c + \hat{H}' + \hat{H}''+... \,,
\end{align}
where $\hat{H}_c$ is the full Hamiltonian evaluated at the centre-of-mass position, and $\hat{H}'$ ($\hat{H}''$) are first-(second-)order gradient corrections in the electromagnetic fields. 

In this semi-classical treatment, we assume that the external fields are sufficiently weak that motion is adiabatic and the wave packet remains in a single non-degenerate energy band at all times. For example, when the electromagnetic perturbations are very small, the assumption of adiabaticity can be made for a wavepacket constructed out of the eigenstates of $\hat{H}_c$ \cite{chang1, chang2}. These states correspond to the original (magnetic) Bloch states $| {u}_n \rangle $ [up to an unimportant phase factor proportional to ${\bs A}({\bs r}_c)$] \cite{chang2}. In this case, the resulting semi-classical equations capture all effects up to first-order in the external perturbing fields. 

Here, we must go a step further as 4D quantum Hall physics is nonlinear in the external electric and magnetic fields, and so we require a semi-classical treatment that is valid up to second order in the perturbations. The wave packet should then instead be constructed out of the perturbed states \cite{XiaoPol, Gao, Gao2}
\begin{align}
| \tilde{u}_n \rangle = | {u}_n \rangle + | \tilde{u}_n' \rangle  \,,
\end{align}
where the eigenstates $ | \tilde{u}_n' \rangle$ are the first-order band-mixing corrections from the perturbing term $\hat{H}'$. With respect to these perturbed states, the semi-classical motion is adiabatic even up to second-order in the external fields, and we can assume that the wave packet remains at all times within the perturbed band $n$. Hereafter, we drop the band index $n$ as well as the subscript $c$. 

Before proceeding, it is important to note that the geometrical properties of the perturbed Bloch band are encoded in a modified Berry curvature \cite{Gao}
\begin{align}
\tilde{\Omega}^{\mu \gamma} &= \Omega^{\mu \gamma} +  \Omega_{1}^{\mu \gamma} \,, \nonumber \\
\Omega^{\mu \gamma} &=  \partial_{k_\mu} \mathcal{A}_{k_\gamma }  -  \partial_{k_\gamma} \mathcal{A}_{k_\mu} \,, \nonumber \\
 \Omega_1^{\mu \gamma} &=  \partial_{k_\mu} \mathcal{A}'_{k_\gamma }  -  \partial_{k_\gamma} \mathcal{A}'_{k_\mu} \,, \label{eq:berry}
\end{align}
where $\mathcal{A}_{k_\mu} = i \langle u |\partial_{k_\mu }| u \rangle $ is the unperturbed Berry connection, containing the effects of any {\it strong} magnetic fields incorporated into the original Bloch band, while $\mathcal{A}'_{k_\mu } = i \langle u |  \partial_{k_\mu } | u' \rangle + \mbox{c.c.}$ is the first-order correction from the {\it perturbing} electromagnetic fields. We note that the correction $\mathcal{A}'_{k_\mu }$ is gauge-invariant, and corresponds to a physical shift of the wave packet centre induced by interband mixing from the external fields \cite{Gao}. 

As derived in Ref.~\cite{Gao}, the semi-classical equations in dimensions $d \ge 2$, valid to second-order in the external fields, are
\begin{align}
\dot{r}^\mu ({\bs k}) &= \frac{\partial \tilde{\mathcal{E}} (\bs k)}{\partial k_\mu } - \dot{k}_\nu \tilde{\Omega}^{\mu \nu}  (\bs k), \label{eq:semir} \\
\dot{k}_\mu &= - E_\mu - \dot{r}^\nu  B_{\mu \nu} \,, \label{eq:semik}
\end{align}
where $\hbar =1$. Here, we have also introduced the second-order wave packet energy, $\tilde{\mathcal{E}}({\bs k})$, which contains the unperturbed Bloch band $\mathcal{E}({\bs k})$ plus corrections from the perturbing terms $\hat{H}' + \hat{H}''$~\cite{Gao, Gao2}.

The above equations of motion can be compared to those from usual first-order semi-classics \cite{chang1, chang2}. In this latter case, the force equation (\ref{eq:semik}) is unchanged, while the velocity equation (\ref{eq:semir}) is simplified: $\tilde{\Omega}^{\mu \gamma} ({\bs k})$ reduces to $\Omega^{\mu \gamma}({\bs k}) $, and $\tilde{\mathcal{E}} ({\bs k})$ reduces to $\mathcal{E}({\bs k})$ plus the correction from $\hat{H}'$. As we now show, none of the corrections to the band energy nor the Berry curvature affect 4D quantum Hall transport for a filled band; hence, for simplicity of presentation, we have not discussed these in the main text. They would, however, have important consequences, for example, in the anomalous Hall physics of partially-filled bands, as previously studied in lower dimensions \cite{Xiao_rev, Gao, Gao2}.

To proceed, we repeatedly combine (\ref{eq:semir}) and (\ref{eq:semik}) to find the mean velocity:
\begin{align}
\dot{r}^\mu &= \frac{\partial \tilde{\mathcal{E}}}{\partial k_\mu }+  E_\nu \tilde{\Omega}^{\mu \nu}+  \dot{r}^\gamma  B_{ \nu \gamma} \tilde{\Omega}^{\mu \nu} \label{eq:meanv}\\
&=\resizebox{.85\hsize}{!}{$\displaystyle{ \frac{\partial \tilde{\mathcal{E}}}{\partial k_\mu }+  E_\nu  \tilde{\Omega}^{\mu \nu}+ \left( \frac{\partial\tilde{\mathcal{E}}}{\partial k_\gamma }  +  E_\delta  \tilde{\Omega}^{\gamma \delta} +  \dot{r}^\alpha  B_{ \delta \alpha}  \tilde{\Omega}^{\gamma \delta} 
\right)B_{ \nu \gamma}  \tilde{\Omega}^{\mu \nu}}$}
\nonumber \\
&\approx \resizebox{.85\hsize}{!}{$\displaystyle{\frac{\partial \tilde{\mathcal{E}}}{\partial k_\mu }+  E_\nu  \tilde{\Omega}^{\mu \nu}+ \left( \frac{\partial\tilde{\mathcal{E}}}{\partial k_\gamma }  +  E_\delta  \tilde{\Omega}^{\gamma \delta} +   \frac{\partial \tilde{\mathcal{E}}}{\partial k_\alpha }   B_{ \delta \alpha}  \tilde{\Omega}^{\gamma \delta} 
\right)B_{ \nu \gamma}  \tilde{\Omega}^{\mu \nu}}$} \,,\nonumber
\end{align}
where in the last line, we have neglected terms that are above second-order in the perturbing electromagnetic fields. We note this approach is similar to that in Ref.~\cite{XiaoPol} which derived the contribution from the $\Snd$ Chern form to the polarisation of an inhomogeneous crystal. 

To calculate the current density of a filled band, we sum the mean velocity over all occupied momentum states and divide by the volume of the system. In the semi-classical limit in 4D, the summation is replaced by
\begin{align}
\frac{1}{V} \sum_{\bs k} \rightarrow \int_{\mathbb{T}^4} \text{d}^4 k D({\bs r}, {\bs k}) \,,\label{eq:sum}
\end{align}
where $V$ is the volume of the system and where the integration is performed over the first (magnetic) Brillouin zone ($\mathbb{T}^4$). Here, we have introduced $D({\bs r}, {\bs k}) $, the modified phase-space density of states, which we shall discuss in more detail below (see Sec.~\ref{sec:phase}). We have generalised this function from lower dimensions \cite{Xiao_ED, Duval, Bliokh, Gosselin, Xiao_reply}, finding in 4D%
\begin{align}
D({\bs r}, {\bs k}) = \frac{1}{(2\pi)^4} & \left[  1 + \frac{1}{2}  B_{\mu \nu} \tilde{\Omega}^{\mu \nu}  \right. +\frac{1}{64} \left( \varepsilon^{\alpha \beta \gamma \delta } B_{\alpha \beta }B_{\gamma \delta }\right)  \nonumber \\ 
& \left. \times \left( \varepsilon_{\mu \nu \lambda \rho} {\tilde{\Omega}}^{\mu \nu }\tilde{\Omega}^{\lambda \rho }\right) \right]  \,, \label{eq:dos}
\end{align}
where all indices now take the values $\mu\!=\!x,y,z,w$.

Combining the above equations and keeping all terms up to second order we find the current density 
\begin{widetext}
\begin{align}
j^\mu =& \int_{\mathbb{T}^4} { \text{d}^4 k } [ \dot{{ r}}^\mu D({\bs r}, {\bs k})]  \label{eq:intcurrent} \\
 \approx& \int_{\mathbb{T}^4} \frac{ \text{d}^4 k }{(2\pi)^4} 
  \left[ \frac{\partial \tilde{\mathcal{E}}}{\partial k_\mu }+  E_\nu  \tilde{\Omega}^{\mu \nu} +
\left(  E_\delta \Omega^{\gamma \delta}B_{ \nu \gamma} \Omega^{\mu \nu} +  \frac{1}{2}E_\nu  {\Omega}^{\mu \nu} B_{\delta \gamma}  {\Omega}^{\delta \gamma}\right) 
+ 
  \left(  \frac{\partial \tilde{\mathcal{E}}}{\partial k_\gamma }  B_{ \nu \gamma} \tilde{\Omega}^{\mu \nu} + \frac{1}{2} \frac{\partial \tilde{\mathcal{E}}}{\partial k_\mu }B_{\gamma \nu}  \tilde{\Omega}^{\gamma \nu} \right) 
   \right.
  \nonumber \\ 
 &\left.+ \left( \left( \frac{\partial \tilde{\mathcal{E}}}{\partial k_\alpha } B_{ \delta \alpha} \Omega^{\gamma \delta} 
+ \frac{1}{2}\frac{\partial \tilde{\mathcal{E}}}{\partial k_\gamma }  B_{\delta \alpha}  {\Omega}^{\delta \alpha}  \right) B_{ \nu \gamma} \Omega^{\mu \nu}+ 
  \frac{1}{64}  \frac{\partial \tilde{\mathcal{E}}}{\partial k_\mu }( \varepsilon^{\alpha \beta \gamma \delta } B_{\alpha \beta }B_{\gamma \delta} )( \varepsilon_{\xi \nu \lambda \rho} {{\Omega}}^{\xi \nu }{\Omega}^{ \lambda \rho} )   \right) 
  \right] \,,\label{eq:current}
 \end{align}
 \end{widetext}
where in the second line we have expanded and grouped together terms of a similar form. In this expression, we have also replaced the modified Berry curvature $ \tilde{\Omega}$ with the unperturbed Berry curvature $\Omega$ wherever appropriate to ensure that we only keep terms up to second-order in the external fields. 

Despite the seeming complexity of (\ref{eq:current}), it is possible to simplify the current density down to Eq. (5) in the main text. We shall now briefly summarise the key steps that are required, taking groups of terms in turn, beginning with the first two:
\begin{align}
\int_{\mathbb{T}^4} \frac{ \text{d}^4 k }{(2\pi)^4} \left( \frac{\partial \tilde{\mathcal{E}}}{\partial k_\mu }+  E_\nu  \tilde{\Omega}^{\mu \nu}  \right)\,.
\end{align}
As can be seen, these terms are very similar to those found for a 2D quantum Hall system within usual first-order semi-classics~\cite{Xiao_rev, Price:2012, Dauphin:2013}. The key differences here are the integral over a 4D rather than a 2D Brillouin zone, as well as the corrected energy $\tilde{\mathcal{E}} ({\bs k})$ and the modified Berry curvature $\tilde{\Omega} ({\bs k})$. However, it can be shown that both the corrected group velocity $\partial_{k_\mu}\tilde{\mathcal{E}}$ and the Berry curvature correction $\Omega_1 ({\bs k})$ vanish upon integration over the (magnetic) Brillouin zone \cite{Gao}. The remaining contribution to the 4D current density is then:
\begin{align}
 E_\nu\int_{\mathbb{T}^4} \frac{ \text{d}^4 k }{(2\pi)^4}  {\Omega}^{\mu \nu}\,,
\end{align}
which is closely related to the 2D quantum Hall (linear) response as discussed in the main text. 

The second set of terms in the current density (\ref{eq:current}) are those proportional to both the external electric and magnetic fields:
\begin{align}
\int_{\mathbb{T}^4} \frac{ \text{d}^4 k }{(2\pi)^4}  
\left(   E_\delta \Omega^{\gamma \delta}B_{ \nu \gamma} \Omega^{\mu \nu} +  \frac{1}{2}E_\nu  {\Omega}^{\mu \nu} B_{\delta \gamma}  {\Omega}^{\delta \gamma}\right) \,.
\end{align}
This can be simplified by using the antisymmetry of the magnetic field strength ($B_{ \gamma \nu}= - B_{ \nu \gamma}$) and the Berry curvature ($\Omega^{ \gamma \nu}=-\Omega^{ \nu \gamma}$). By writing out all terms and cancelling those which vanish by symmetry, it can be shown that this set of terms reduce to:
\begin{align}
 \frac{\nu_2}{(2  \pi)^2} \varepsilon^{\mu \alpha \beta \nu} E_\nu B_{ \alpha \beta} \,,
 \end{align}
where we have introduced the second Chern number for a spinless system~\cite{Ryu}:
\begin{align}
\nu_2\!&=\!\frac{1}{8 \pi^2} \int_{\mathbb{T}^4}  \Omega \wedge \Omega   \in  \mathbb{Z}\,, \notag \\
&=\!\frac{1}{4 \pi^2} \int_{\mathbb{T}^4}  \Omega^{xy}\Omega^{zw}\!+\! \Omega^{wx}\Omega^{zy}\!+\! \Omega^{zx}\Omega^{yw} \text{d}^4k\,.\label{second_chern}
\end{align} 
In the second line, we have written out explicitly the sum over the four dimensions for clarity. 

The third set of terms in the current density (\ref{eq:current}) are first-order combinations of the external magnetic field, the Berry curvature and the group velocity:
\begin{align}
\int_{\mathbb{T}^4} \frac{ \text{d}^4 k }{(2\pi)^4}    \left(  \frac{\partial \tilde{\mathcal{E}}}{\partial k_\gamma }  B_{ \nu \gamma} \tilde{\Omega}^{\mu \nu} + \frac{1}{2} \frac{\partial \tilde{\mathcal{E}}}{\partial k_\mu }B_{\gamma \nu}  \tilde{\Omega}^{\gamma \nu} \right)  \,.
  \end{align}
As we have assumed summation over all repeated indices, this expression contains many terms; we now group these into sets that contain each antisymmetric pair of magnetic field strength components. For example, collecting all terms with either $B_{ x y}$ or $B_{  y x}(=- B_{   x y})$, we find: 
\begin{align}  B_{x y }   \int_{\mathbb{T}^4} \frac{ \text{d}^4 k }{(2\pi)^4} \left( 
\frac{\partial  \tilde{\mathcal{E}}}{\partial k_\mu } \tilde{\Omega}^{x y}+   \frac{\partial  \tilde{\mathcal{E}}}{\partial k_x }  \tilde{\Omega}^{ y \mu}+  \frac{\partial  \tilde{\mathcal{E}}}{\partial k_y }   \tilde{\Omega}^{\mu x} \right)\,,
\end{align}
where we have used the antisymmetry of the modified Berry curvature. As both $\tilde{\mathcal{E}} ({\bs k})$ and $\tilde{\Omega} ({\bs k})$ are periodic over the  Brillouin zone, we can integrate by parts to find:
\begin{align}  
B_{x y }   \int_{\mathbb{T}^4} \frac{ \text{d}^4 k }{(2\pi)^4}  
\tilde{\mathcal{E}} \left( 
\frac{\partial \tilde{\Omega}^{x y }}{\partial k_\mu }+  \frac{\partial  \tilde{\Omega}^{ y \mu}}{\partial k_x } +  \frac{\partial \tilde{\Omega}^{\mu x}}{\partial k_y}    \right) \,,
\end{align}
which vanishes due to the Bianchi identity \cite{XiaoPol}:
\begin{align}
\frac{\partial \tilde{\Omega}^{x y }}{\partial k_\mu }+  \frac{\partial  \tilde{\Omega}^{ y \mu}}{\partial k_x } +  \frac{\partial \tilde{\Omega}^{\mu x}}{\partial k_y}  =0\,.
\end{align}
The same argument can be repeated for all other pairs of magnetic field strength components. As a result, this third set of terms makes no overall contribution to the current of a filled band. 

The final set of terms in the current density (\ref{eq:current}) are higher-order combinations of the external magnetic field, the Berry curvature and the group velocity:
\begin{align}
 \int_{\mathbb{T}^4} &\frac{ \text{d}^4 k }{(2\pi)^4} 
  \left[ \left( \frac{\partial \tilde{\mathcal{E}}}{\partial k_\alpha } B_{ \delta \alpha} \Omega^{\gamma \delta} 
+ \frac{1}{2}\frac{\partial \tilde{\mathcal{E}}}{\partial k_\gamma }  B_{\delta \alpha}  {\Omega}^{\delta \alpha}  \right) B_{ \nu \gamma} \Omega^{\mu \nu}\right. \nonumber \\
& + \left.
  \frac{1}{64}  \frac{\partial \tilde{\mathcal{E}}}{\partial k_\mu }( \varepsilon^{\alpha \beta \gamma \delta } B_{\alpha \beta }B_{\gamma \delta} )( \varepsilon_{\xi \nu \lambda \rho} {{\Omega}}^{\xi \nu }{\Omega}^{ \lambda \rho} )   \right]  \,. 
 \end{align}
Using only the antisymmetry of the magnetic field strength and of the Berry curvature, it can be shown that this set of terms sums to zero, giving no contribution to the integral. We note that this cancellation is by symmetry alone, and so holds also for the current density of a band with arbitrary filling.    

Combining all these steps, the current density (\ref{eq:current}) of a filled band up to second-order in the external fields can be simplified to: 
\begin{align}
j^\mu
   &= E_\nu  \frac{1}{(2 \pi)^4} \int_{\mathbb{T}^4}
 {\Omega}^{\mu \nu}   \text{d}^4 k+   \frac{\mathcal{\nu}_2}{4 \pi^2} \varepsilon^{\mu \alpha \beta  \nu}  E_\nu B_{ \alpha \beta} \,,\label{eq:main}
 \end{align}
as stated in the main text. As can be seen, this expression does not depend on any of the corrections to the energy and Berry curvature that appeared in the full second-order semi-classical equations (\ref{eq:semir}) and (\ref{eq:semik}).

\section{The Modified Phase-Space Density of States $D({\bs r}, {\bs k})$ in 4D} \label{sec:phase}
\setcounter{enumi}{2}
\setcounter{equation}{0}

As discussed above and in the main text, to calculate the transport equation for a filled band in the semi-classical limit [Eq.~(\ref{eq:intcurrent})], we integrate the mean velocity over the Brillouin zone, taking into account a modified phase-space density of states \cite{Xiao_ED, Duval, Bliokh, Gosselin}. We now discuss the interpretation of this factor and derive its extension [Eq.~(\ref{eq:dos})] to four dimensions. 

In the absence of the Berry curvature and/or the external ${\bs B}$ field, the phase-space density of states is a constant factor, given by $1/ (2 \pi)^d$. Classically, the independence of the density of states from the phase-space coordinates is a consequence of Liouville's theorem, which states that the phase-space volume element is conserved under time evolution \cite{Xiao_ED}. However, when both the magnetic field and Berry curvature are non-zero, it can be shown that Liouville's theorem applies to the volume element $\Delta V = \Delta {\bs R} \Delta {\bs K}$ of the {\it canonical} position ${\bs R}$ and momentum ${\bs K}$, but not necessarily to the volume element $\Delta V' = \Delta {\bs r} \Delta {\bs k}$ of the {\it physical} position ${\bs r}$ and momentum ${\bs k}$ \cite{Xiao_ED, Bliokh, Duval}. This is because the relationship between the canonical and physical variables is, in general, strongly affected by the presence of the magnetic vector potential and the Berry connection. 

In the simplest case, when we have a magnetic field but no Berry curvature, it is well-known that the physical momentum is modified by the Peierls substitution ${\bs k} = {\bs K} - {\bs A} ({\bs r})$, while the physical position remains equal to the canonical position \cite{Landau}. If instead, we have Berry curvature but no magnetic field, the physical position becomes ${\bs r} = {\bs R} + { \mathcal{A}} ({\bs k})$, while the physical and canonical momenta remain the same \cite{Adams, Nagaosa, Xiao_rev}. This is because the Berry curvature acts as an artificial magnetic-field in momentum space \cite{Nagaosa, Bliokh_mag}. Consequently, in the presence of both Berry curvature and a magnetic field, neither the physical position nor momentum remain equal to their canonical counterparts. Instead, we must apply generalised Peierls substitutions for both variables which contain many additional nontrivial terms; this is discussed in detail in Refs.~\cite{Gao, Xiao_rev, Bliokh, Gosselin}.

Returning to Eq.~(\ref{eq:sum}), we see that the violation of Liouville's theorem for the physical variables means that, in general, the phase-space density of states with respect to ${\bs r}$ and ${\bs k}$ is not a constant. Hence, when we replace the sum over physical momenta by an integral, we include the factor $D({\bs r}, {\bs k})$ which can depend on both the physical position and momentum. We emphasise again that while the violation of Liouville's theorem may at first seem surprising, the dynamics are of course still Hamiltonian because of the preservation of Liouville's theorem for the canonical variables \cite{ Bliokh, Duval}.

To understand $D({\bs r}, {\bs k})$ further, there are several equivalent approaches that have been discussed in the literature \cite{Xiao_ED, Duval, Bliokh, Gosselin, Gosselin2}. Firstly, as in Ref.~\cite{Xiao_ED}, we could consider the time evolution of the physical volume element given by: 
\begin{align}
\frac{1}{\Delta V'} \frac{\partial \Delta V'}{\partial t} = \frac{\partial}{\partial r^\mu } \dot{r}^\mu + \frac{\partial}{\partial k_\mu } \dot{k}_\mu \,. \label{eq:time}
\end{align}
Substituting in from Eq.~(\ref{eq:semir}) and Eq.~(\ref{eq:semik}), this can be solved analytically for $\Delta V'({\bs r}, {\bs k})$. The modified density of states is then chosen to ensure that $D({\bs r}, {\bs k})
\Delta V'({\bs r}, {\bs k})$ remains constant at all times, i.e., that the change of the volume element under time evolution is compensated for by the change in the density of states \cite{Xiao_ED}. In this sense, it is natural to call $D({\bs r}, {\bs k})$ the modified phase-space density of states.

A second interpretation of $D({\bs r}, {\bs k})$ is as the Jacobian of the transformation from the canonical to physical variables multiplied by a factor $1/ (2 \pi)^d$ \cite{Duval, Bliokh, Gosselin, Gosselin2}. In this picture, we see that the integral of the velocity in Eq.~(\ref{eq:intcurrent}) is constructed first for the canonical variables, where Liouville's theorem applies and the phase-space density of states is $1/ (2 \pi)^d$. Then the integral is transformed by the Jacobian into the physical variables. In the absence of either a magnetic field or the Berry curvature, it can be shown that the Jacobian is equal to unity and the transformation is unimportant \cite{Duval, Xiao_reply}. 

This Jacobian description also leads to simple derivation of $D({\bs r}, {\bs k})$ from classical Hamiltonian dynamics~\cite{Duval}. As this is the most straightforward calculation for $D({\bs r}, {\bs k})$, we apply this method to find the modified 4D phase-space density of states for Eqs.~(\ref{eq:semir}) and (\ref{eq:semik}). In classical Hamiltonian dynamics, the Euler-Lagrange equations can be written as :
\begin{align}
\omega_{\alpha \beta} \dot{\xi}^\beta &= \frac{\partial h } {\partial \xi^\alpha}\,, 
\end{align}
where $h$ is the classical Hamiltonian, $\xi$ denotes the phase-space physical coordinates ${\bs r}$ and ${\bs k}$ collectively, and $\omega_{\alpha \beta}$ is a symplectic matrix. Re-writing Eqs.~(\ref{eq:semir}) and (\ref{eq:semik}) in this form, we have in full:
\begin{widetext}
\begin{align}
 \left(
\begin{array} {cccccccc}
 0 & - B_{x y} & - B_{x z} & - B_{x w} & -1 & 0 & 0 & 0\\
 B_{x y } & 0 & - B_{y z} & - B_{y w} & 0 & -1 & 0 & 0\\
  B_{x z } &    B_{y z} &0 & - B_{z w}& 0 & 0 & -1 & 0\\
  B_{x w } &    B_{y w} &B_{z w} & 0 & 0 & 0 & 0 & -1\\
  1 &    0&0 & 0 &0  &  \tilde{\Omega}^{x y} &  \tilde{\Omega}^{x z}&  \tilde{\Omega}^{x w}\\  
  0 &    1&0 &0 & -  \tilde{\Omega}^{x y}&0 &  \tilde{\Omega}^{y z}&  \tilde{\Omega}^{y w}\\    
  0 &    0&1 &0 & -  \tilde{\Omega}^{x z}&  -  \tilde{\Omega}^{ y z} & 0 &   \tilde{\Omega}^{z w}\\
    0 &    0&0 &1 & -  \tilde{\Omega}^{x w}&  -  \tilde{\Omega}^{ y w} & -  \tilde{\Omega}^{z w} &  0\\    
   \end{array}
\right)
 \left(
\begin{array} {c}
\dot{x} \\
\dot{y} \\
\dot{z} \\
\dot{w} \\
\dot{k}_x \\
\dot{k}_y \\
\dot{k}_z \\
\dot{k}_w 
\end{array}
\right)
=
 \left(
\begin{array} {c}
E_x \\
E_y \\
E_z \\
E_w \\
\frac{\partial \tilde{\mathcal{E}}}{\partial k_x }\\
\frac{\partial \tilde{\mathcal{E}}}{\partial k_y } \\
\frac{\partial \tilde{\mathcal{E}}}{\partial k_z } \\
\frac{\partial \tilde{\mathcal{E}}}{\partial k_w }
\end{array}
\right)\,,
\end{align} 
\end{widetext}
where we identify the classical Hamiltonian as $h=\tilde{\mathcal{E}} + E_\mu r^\mu$. Then, it can be shown that the Jacobian of the transformation from the canonical to  physical variables can be simply calculated as the square-root of the determinant of the symplectic matrix \cite{Duval}. In 4D, this is:
\begin{align}
D({\bs r}, {\bs k})= \frac{1}{(2 \pi)^4} &\sqrt{\mbox{det}(\omega)}  \\ =   \frac{1}{(2 \pi)^4} &\left[  1 + \frac{1}{2}  B_{\mu \nu} \tilde{\Omega}^{\mu \nu}  \right. +\frac{1}{64} \left( \varepsilon^{\alpha \beta \gamma \delta } B_{\alpha \beta }B_{\gamma \delta }\right)  \nonumber \\ 
& \left. \times \left( \varepsilon_{\mu \nu \lambda \rho} {\tilde{\Omega}}^{\mu \nu }\tilde{\Omega}^{\lambda \rho }\right) \right]\,,\nonumber 
\end{align}
as stated above and in the main text. In less than four dimensions, the last term vanishes due to the Levi-Civita symbols, and what remains is 
\begin{align}
D_{\text{2D/3D}}({\bs r}, {\bs k})&=  1 + \frac{1}{2} B_{\mu \nu} \tilde{\Omega}^{\mu \nu} \,,
\end{align}
which is the previously known result for two and three dimensions \cite{Xiao_ED, Duval}. 

Finally, we note that while the above discussion was framed in terms of classical variables, these issues can also be approached from a quantum perspective, for example by re-quantizing the semi-classical equations \cite{Xiao_ED} or by beginning from the full quantum Hamiltonian and performing a semi-classical diagonalization \cite{Gosselin, Gosselin2}. 

\section{The Streda-Widom Formula} \label{sec:streda}
\setcounter{enumi}{3}
\setcounter{equation}{0}
The Streda-Widom formula is a simple thermodynamic relation to calculate the quantum Hall response of an insulating material \cite{Widom, Streda, Smrcka}. While this was originally derived for the linear 2D quantum Hall effect, it has since been discussed in the context of nonlinear physics in Ref.~\cite{Bergman}, where the magnetoelectric coupling of a three-dimensional topological insulator was studied. We now show that a direct application of the Streda-Widom formula appears to also capture the physics of 4D quantum Hall transport.  

The bulk of a system is insulating provided that the chemical potential lies in a band gap. Under the application of an electric field, local equilibrium is established with the electric field $E_\nu$ imposing a gradient of the chemical potential $\mu$ across the system \cite{Smrcka}. If the electric field is sufficiently weak, the bulk remains insulating as the chemical potential stays in the gap. However, the boundaries of the system can support surface currents, which are a function of the magnetic moment evaluated at the local chemical potential \cite{Streda, Smrcka}. From thermodynamics, it can be shown that the current density of the surface currents is given by \cite{Smrcka}:
\begin{align}
 j^\alpha &=  -    \left(\frac{\partial M^{\nu \alpha}}{
\partial \mu}\right)_{T, B} {E}_\nu\nonumber\\
&=-    \left(\frac{\partial n}{
\partial B_{ \nu \alpha}}\right)_{T, \mu} {E}_\nu\,,
\end{align}
where $M^{\nu \alpha}$ is the magnetic moment, $n$ is the particle density and $T$ is the temperature. In 2D, it has been shown that this current density may be quantised in units of the first Chern number; this is the standard quantum Hall effect \cite{Widom, Streda, Smrcka, Xiao_ED}.

We now apply this formula in four dimensions; then the density of particles is
\begin{align}
n=  \int_{\mathbb{T}^4}  \text{d}^4 k D ({\bs r}, {\bs k}) \Theta (\varepsilon_F - \varepsilon)\,, \label{eq:dens}
\end{align}
as extended from lower dimensions \cite{Xiao_ED}. Here, $\Theta (\varepsilon_F - \varepsilon)$ is the distribution function of fermions, which uniformly fill the band up to Fermi energy $\varepsilon_F$. We assume that the Fermi energy lies in a band-gap, such that the band is completely filled. 

As can be seen from (\ref{eq:dens}), the only dependence of the particle density on a perturbing magnetic field, at a constant chemical potential, comes from the modified phase-space density of states $D ({\bs r}, {\bs k})$ \cite{Xiao_ED}. Then considering, for example, the surface current density along the $x$ direction under an applied electric field $E_y$, it is straightforward to show that:
\begin{align}
j^x &= -  \left(\frac{\partial n}{
\partial B_{y x}}\right)_{T, \mu} {E}_y
\nonumber \\
&=  E_y \int_{\mathbb{T}^4}   \frac{\text{d}^4 k}{(2 \pi)^4} {\Omega}^{xy}  + E_y  B_{zw} \frac{\mathcal{\nu}_2 }{(2 \pi)^2} \,,
\end{align}
where we have used that the corrections $ \Omega_1$ vanish upon integration over the whole Brillouin zone, and have only retained terms up to second-order in the fields. We have also applied the definition of the second Chern number for a spinless system (\ref{second_chern}). This procedure can be repeated for all other configurations, recapturing the general result for the 4D current density stated above (\ref{eq:main}); this suggests the central role of the modified 4D phase-space density of states in the physics of the 4D quantum Hall effect. 

\section{Spectral properties and Chern number decomposition}
\label{sec:spec}
\setcounter{enumi}{4}
\setcounter{equation}{0}
In this Appendix, we elaborate on the spectral details of the model studied in the main text [cf.~Eq.~(7)]. 
We are considering a 4D model that is a direct sum of two copies of the 2D Hofstadter model \cite{Hofstadter}. For emphasis, let us rewrite Eq.~(7) in another form
\begin{align}
\hat H=&- J \sum_{w,x,y,z} \Big(\hat h^{\rm Hofstadter}_{w,y}(x,z,\Phi_1)\\
& \qquad\qquad\qquad\qquad + \hat h^{\rm Hofstadter}_{x,z}(y,w,\Phi_2)\Big)\,,\nonumber
\end{align}
where 
\begin{align}
\hat h^{\rm Hofstadter}_{\alpha,\beta}(\gamma,\delta,\Phi) = &c^{\dagger}_{\alpha,\beta,\gamma+1,\delta} c^{\phantom{{\dagger}}}_{\alpha,\beta,\gamma,\delta}\\
& + e^{i 2 \pi \Phi \gamma}c^{\dagger}_{\alpha,\beta,\gamma,\delta+1}c^{\phantom{{\dagger}}}_{\alpha,\beta,\gamma,\delta} + \text{h.c.}\,,\nonumber
\end{align}
and here we have assumed that the lattice spacing $a=1$. 

\begin{figure}[tbh]
\begin{center}
\includegraphics[width=\columnwidth]{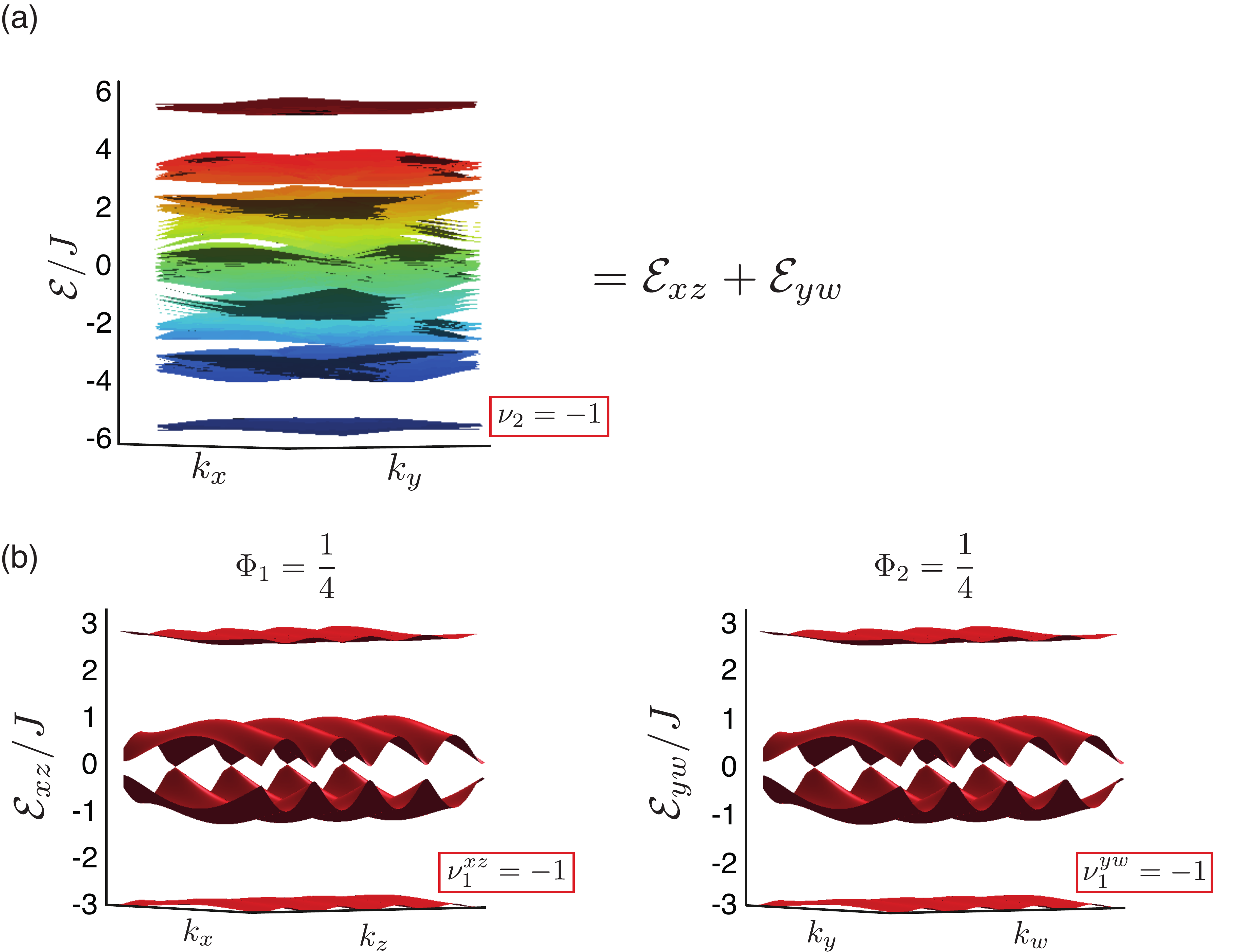}
\end{center}
\caption{  \label{hofstadter_spec} %
(a) The energy spectrum $\mathcal{E}(k_x,k_y)$, as shown in the main text, of the 4D model for $\Phi_{1,2}\!=\!1/4$, and for many values $k_{z,w}$ over the Brillouin zone. The 2CN of the lowest band is indicated. This 4D spectrum is a Minkowski sum of the spectra $\mathcal{E}_{xz} ({\bs k})$ and $\mathcal{E}_{yw} ({\bs k})$  of the 2D Hofstadter models in the $x-z$ [resp. $y-w$] plane, shown in (b) on the LHS [resp. RHS]. The first Chern numbers of the lowest band for each 2D Hofstadter model is indicated.}
\end{figure}

The spectrum of each Hofstadter model is obtained by solving the 2D Harper equations 
\begin{widetext}
\begin{align}
\mathcal{E}_{xz}(k_x, k_z) \phi_x (k_x, k_z)=&-J \Biggl \{ e^{i k_x} \phi_{x+1} (k_x, k_z) +e^{-i k_x} \phi_{x-1} (k_x, k_z)  - 2 \phi_x (k_x, k_z) \cos (2 \pi \Phi_1 x +k_z)\Biggr \} \,,\\
\mathcal{E}_{yw} (k_y, k_w) \chi_y (k_y, k_w)=&-J \Biggl \{ e^{i k_y} \chi_{y+1}(k_y, k_w) +e^{-i k_y} \chi_{y-1}(k_y, k_w)   - 2 \chi_y (k_y, k_w)  \cos (2 \pi \Phi_2 y +k_w)\Biggr \} \, ,
\end{align}
\end{widetext}
where $x\!=\!\{1, \dots, q_1\}$, $y\!=\{\!1, \dots, q_2\}$ for $\Phi_1\!=\!p_1/q_1$ and  $\Phi_2\!=\!p_2/q_2$, and where $p_{1,2}, q_{1,2}$ are integers; here the eigenfunctions satisfy the periodicity conditions $\phi_x (k_x, k_z)=\phi_{x+q_1} (k_x, k_z)$ and $\chi_y (k_y, k_w)  =\chi_{y+q_2} (k_y, k_w) $. For example, the spectrum of each 2D Hofstadter model with flux $\Phi=1/4$ is seen in Fig.~\ref{hofstadter_spec}(b).

The energy spectrum of the full 4D model, as seen in the main text and here in Fig.~\ref{hofstadter_spec}(a), is a Minkowski sum of its two Hofstadter constituents, i.e., 
\begin{align}
\mathcal{E}({\bs k})\!=\!\{ E_1 +E_2 \vert E_1\in \mathcal{E}_{xz}(k_x, k_z), E_2 \in \mathcal{E}_{yw}(k_y, k_w) \} \nonumber.
\end{align}
As a result, filling up the lowest band, up to a chemical potential $\mu$ within the first gap of $\mathcal{E}({\bs k})$, corresponds to a Minkowski sum over the energies of the lowest band of each Hofstadter constituent. Thus, when evaluating the $\Snd$ Chern number, $\nu_2$ [cf.~Eq.~(6) in the main text], the integral can be split such that each 2D Berry curvature is integrated up to the lowest band of the corresponding Hofstadter spectrum, and we obtain the result appearing in the main text $\nu_2\!=\!\nu_1^{zx}\! \times \! \nu_1^{yw}$, where we introduced the $\Fst$ Chern numbers 
\be
\nu_1^{\mu \nu}= \frac{1}{2\pi} \int_{\mathbb{T}^2}  \, \Omega^{\mu \nu} (k_{\mu}, k_{\nu}) \, \text{d} k_{\mu} \text{d} k_{\nu} ,
\ee
associated with the lowest bands of the 2D (Hofstadter) spectra $\mathcal{E}_{xz}$ and $\mathcal{E}_{yw}$, where the integral is over the 2D (magnetic) Brillouin zone ($\mathbb{T}^2$) for momenta $k_{\mu}, k_{\nu}$.

\section{Numerical evaluation of the Berry curvature and 2nd Chern number}
\setcounter{enumi}{5}
\setcounter{equation}{0}
In this Appendix, we briefly describe a numerical method allowing for the computation of the Berry curvature $\Omega^{\mu \nu} (\bs k)$ in 4D lattice-systems, i.e. $\mu, \nu=x,y,z,w$. In particular, this efficient method offers a powerful method to numerically evaluate the $\Snd$ Chern number of 4D Bloch bands, defined in Eq. \eqref{second_chern} above. We emphasize that the algorithm described here is a direct generalization of the method developed by Fukui \emph{et al.} \cite{Fukui} for 2D lattice systems; we thus refer the reader to this reference for technical details. 

We start by considering a general 4D lattice system. The momentum-space representation of the lattice Hamiltonian is written as $\hat H (\bs k)$, where $\bs k=\{k_{\mu}\}$ is the quasi-momentum, taking values within the first (magnetic) Brillouin zone. 

For the sake of simplicity, we will consider that the spectrum $\mathcal{E}(\bs k)$ associated with $\hat H (\bs k)$ displays a non-degenerate, well isolated, lowest energy band [e.g. Fig.~\ref{hofstadter_spec}(a)]. Generalization of the method to the case of non-degenerate bands directly follows from the determinant formula [Eq. (16)] in Ref.~\cite{Fukui}. 

The numerical method relies on a discretization of $k$-space: $\hat H (\bs k)$ is solved on a $k$-space lattice, with lattice sites denoted $\bs k_L$. The ground eigenstate $\vert u (\bs k_L) \rangle$ is numerically calculated at each lattice site, which allows us to build a link variable
\be
U^{\mu} (\bs k_L) = \langle u (\bs k_L) \vert u (\bs k_L + \bs e^{\mu}) \rangle / \mathcal{N} \,,
\ee
where $\mathcal{N}$ is a normalization constant. In lattice gauge theory \cite{Fukui}, such link variables are directly related to the curvature through a plaquette product. Here, the (continuous) Berry curvature $\Omega(\bs k)\!=\!(1/2) \Omega^{\mu \nu}(\bs k) \text{d}k_{\mu}\! \wedge \! \text{d}k_{\nu}$ is approximated by \cite{Fukui}
\begin{align}
\Omega^{\mu \nu}(\bs k_L) \!=\! i \ \text{log} \Bigl \{ &U^{\mu} (\bs k_L) U^{\nu} (\bs k_L \!+\! \bs e^{\mu}) \\
&\times \left [U^{\mu} (\bs k_L \!+\! \bs e^{\nu}) \right ]^{-1}   \left [U^{\nu}(\bs k_L)\right ]^{-1}     \Bigr \} \,. \notag
\end{align}
The $\Snd$ Chern number is eventually calculated using the discretized expression
\begin{align}
\nu_2= \frac{1}{4 \pi^2} \sum_{\bs k_L}& \Omega^{xy}(\bs k_L)\Omega^{zw}(\bs k_L) \\
&\!+\! \Omega^{wx}(\bs k_L)\Omega^{zy}(\bs k_L)\!+\! \Omega^{zx}(\bs k_L)\Omega^{yw}(\bs k_L)\,. \notag
\end{align}

We applied this simple algorithm to our model [cf.~Eq.~(7) in the main text], for the flux configuration chosen in the main text. The Hamiltonian matrix $\hat H (\bs k)$ is associated with the (Harper-like) eigenvalue equation
\begin{align}
&\mathcal E(\bs k) u_{x,y} (\bs k)=-J \Biggl \{ e^{i k_x} u_{x+1,y} (\bs k)+e^{-i k_x} u_{x-1,y} (\bs k) \notag \\
&\qquad\qquad\,\, + e^{i k_y} u_{x,y+1} (\bs k)+e^{-i k_y} u_{x,y-1} (\bs k) \\
&\qquad - 2 u_{x,y} (\bs k) \left [\cos (2 \pi \Phi_1 x +k_z) + \cos (2 \pi \Phi_2 y +k_w)\right ]\Biggr \} \notag \label{app_harper} \,,
\end{align}
where $x\!=\!\{1, \dots, q_1\}$, $y\!=\{\!1, \dots, q_2\}$ for $\Phi_1\!=\!p_1/q_1$ and  $\Phi_2\!=\!p_2/q_2$, and where $p_{1,2}, q_{1,2}$ are integers; here again the eigenfunctions satisfy periodicity conditions $u_{x, y}({\bs k})=u_{x+q_1, y} ({\bs k})$ and $u_{x, y}({\bs k})=u_{x, y+q_2}({\bs k})$.

For $\Phi_{1,2}\!=\!\pm1/q$, the $k$-space lattice  was discretized as
\be
k_{\mu} = \frac{2 \pi}{q \mathfrak{N}} n_{\mu}, \, \, \text{ with }  n_{x,y}=1, \dots, \mathfrak{N} \, ;\, \,  n_{z,w}=1, \dots, q\mathfrak{N}\,.\notag
\ee
Stable and accurate results were obtained for $\mathfrak{N} \!\ge\! 3$, namely, for $k$-space lattices of size larger than $3 \times 3 \times 3q \times 3q$. We first verified that the only non-zero components $\Omega^{\mu \nu}(\bs k_L)$  were $\Omega^{zx}(k_x, k_z)$ and $\Omega^{yw}(k_y, k_w)$, which is in agreement with the configuration of the fluxes $\Phi_1$ and $\Phi_2$, which are penetrating the $x\!-\!z$ and $y\!-\!w$ planes, respectively. We also verified that the lowest energy band of the model was indeed associated with a non-zero $\Snd$ Chern number given by $\nu_2\!=\!\nu_1^{zx}\! \times \! \nu_1^{yw}$ [cf.~Sec.~\ref{sec:spec}]. In particular, the numerical method yields a remarkably precise value $\nu_2\!=\! \nu_1^{xz}\!=\! \nu_1^{yw}=-1.$ for $\Phi_{1,2}=1/4$, see Fig.~\ref{hofstadter_spec}, and $\mathfrak{N} \!\ge\! 3$ \cite{Fukui}.  

\section{Partially-filled bands and the center-of-mass displacement}
\setcounter{enumi}{6}
\setcounter{equation}{0}
In this Appendix, we briefly discuss transport for a partially-filled energy band. In this case, the current density in Eq.~\eqref{eq:intcurrent} is replaced by
\be
j^\mu_{\rho} = \int_{\mathbb{T}^4} {\text{d}^4k } \, \rho (\bs k) \, \dot{{ r}}^\mu D(\bs{r}, \bs{k}),\label{app:current_partial}
\ee
where $\rho (\bs k) $ is a filling function, which can be determined by band-mapping techniques~\cite{Aidelsburger:2015}. As in the main text, we still restrict ourselves to the situation where only the lowest band is occupied; generalization to multi-band configurations is straightforward, see Ref.~\cite{Aidelsburger:2015}.

In general, $\rho (\bs k)$ may be an arbitrary function. However, in the experimentally-relevant case where the occupied band has a large flatness ratio $\Delta/W$ [e.g. as in Fig.~\ref{hofstadter_spec}(a)], both the Berry curvature $\Omega^{\mu \nu}(\bs k)\!\approx\!\Omega^{\mu \nu}$ and the filling function $\rho (\bs k)\!\approx\! \rho$ are uniformly distributed over the magnetic Brillouin zone. The latter is because the temperature is generally large compared to the bandwidth, yet still can be made small (or of the order) of the band gap, $W \ll k_{\text{B}}T \ll \Delta$. In fact, the interplay between temperature, inter-particle collisions and the periodic-driving of the lattice can even lead to a uniform population of dispersive bands, as was recently observed in Bose gases loaded into Hofstadter bands~\cite{Aidelsburger:2015}. 

In the following, we consider a single band, filled uniformly with a filling factor $\rho$; the case $\rho=1$ corresponds to that of an exactly filled Bloch band. In this partially-filled-band configuration, the current density in Eq.~(5) is simply modified by the constant filling factor, 
\begin{align}
j^{\mu}_{\rho}\!=\!\rho j^{\mu}\!=\!\rho\, \left \{ E_{\nu} \frac{1}{(2\pi)^4} \int_{\mathbb{T}^4} \Omega^{\mu \nu} \text{d}^4k + \frac{\nu_2}{4 \pi^2} \varepsilon^{\mu \alpha \beta \nu} E_{\nu} B_{\alpha \beta} \right \}, \label{central_transport_equation_part}
\end{align}
which directly follows from a generalization of the calculations in Section I of the Supplemental Material. 

We relate the current density to the centre-of-mass (COM) velocity of a cloud of atoms by noting that the former is given by ${\bs j}_\rho={\bs v}_{\text{tot}}/V$, while the latter is ${\bs v}_\text{c.m.}={\bs v}_{\text{tot}}/N_{\text{tot}}$, for a total particle number $N_{\text{tot}}$ and a total velocity ${\bs v}_{\text{tot}}$. Combining these definitions, we see that ${\bs v}_\text{c.m.}= {\bs j}_\rho/ n$ where $n$ is the particle density. For the configuration of lattice model and perturbing fields considered in the main text, the density of a partially filled band is simply $n=\rho/ V_{\text{cell}}$. Then the COM velocity $v_{\text{c.m.}}^{\mu}$ is 
\begin{eqnarray}
v_{\text{c.m.}}^{\mu} = \frac{v^{\mu}_{\text{tot}}}{N_{\text{tot}}}=\frac{j^{\mu}_{\rho} }{ \rho /V_{\text{cell}}}=j^{\mu}  V_{\text{cell}}. \label{relation_COM_current}
\end{eqnarray}
Hence, unlike the current density, the COM velocity is independent of the filling factor $\rho$. The center-of-mass measurement described in the main text could therefore be equally carried out with systems that only partially fill the lowest (flat) band [e.g. Fig.~\ref{hofstadter_spec}(a)] as with those that completely fill the band. This indicates that the 4D QH physics we have described will be experimentally accessible with both Fermi and Bose gases. 

\section{Open boundary conditions along the synthetic dimension}\label{sect:open}
\setcounter{enumi}{7}
\setcounter{equation}{0}

In this Appendix, we analyse the possibility of measuring the $\Snd$ Chern number without engineering periodic boundary conditions along the synthetic dimension ($w$). 

Indeed, the numerical results shown in Fig. 2 [main text] were obtained in a configuration where $N\!=\!4$ internal states were used to form a \emph{closed} synthetic dimension; from the experimental point-of-view, this necessitates the further coupling of those states delimiting the edges in the $w$ direction (which we loosely denote as $w\!=\!1$ and $w\!=\!N$), see Ref. \cite{Celi:2014}. Here, we instead consider a finite 4D lattice with \emph{open} boundary conditions along all directions, including the synthetic one. We note that current experimental setups using $^{173}$Yb atoms \cite{Mancini:2015} can potentially generate a synthetic dimension with up to $N\!=\!6$ internal states (from the $I=5/2$ nuclear spin manifold). To go further still, the ground-state manifold of $^{40}$K atoms could be used; this has a total angular momentum $F\!=\!9/2$, potentially allowing experiments to work with $N\!\approx\!10$ internal states.

The Chern-number measurement introduced in the main text relies on the detection of the center-of-mass (COM) displacement. This method measures a bulk property, and as such, it remains accurate as long as the edges of the system do not affect the motion of the cloud. In particular, the method breaks down if the cloud suddenly hits the sharp edges at $w\!=\!1$ and $w\!=\!N$. Therefore, even if the system is initially well-localized around the central site $\langle w \rangle \approx N/2$, corresponding to preparing the gas in a small number of internal states, this method would require the use of a large number of internal states $N\!\gg\!1$ to be applicable in practice. 

We now investigate this finite-size effect more quantitatively. Let us consider the relevant setup configuration illustrated in Fig. 1 [main text], for which the COM velocity is given by the expressions 
\begin{align}
&v^x_{\text{c.m.}}=j^x \times 16 a^4\!=\! 8 a^2 E_y \tilde \Phi/\pi , \label{velocities_appendix} \\
&v^w_{\text{c.m.}}=j^w \times 16 a^4\!=\! 2 a^2 E_y / \pi,  \qquad v^y_{\text{c.m.}}=v^z_{\text{c.m.}}=0.\notag 
\end{align}
Here,  we used Eqs. (8)-(9)-\eqref{relation_COM_current}, and have taken into account the fact that $\nu_2\!=\!\nu_1^{yw}\!=\!-1$ for this configuration as the fluxes are $\Phi_{1,2}\!=\!1/4$. In the following, we also set the perturbing flux to $\tilde \Phi=1/10$, and the electric field strength $E_y=0.2 J/a$, as used in the main text [see Fig.~2]. From the COM velocity \eqref{velocities_appendix}, one can readily determine the time $t^*$ at which the cloud touches the edge in the $w$ direction. Let us suppose that the cloud is initially prepared around the central site (with $w$-coordinate $r^w \approx a N/2$), with a spread of about 1-2 lattice sites in the $w$ direction; this corresponds to initially populating 1-2 internal states. Then the critical time is
\be
t^* \approx a \left ( \frac{N}{2} -1 \right ) / v^w_{\text{c.m.}} \approx \frac{5 \pi}{2} \left ( \frac{N}{2} -1 \right ) (\hbar/J),
\ee
where we have reintroduced $\hbar$. This indicates that  the duration of the COM-drift measurement should not exceed the critical time $t^* \!\approx \! 10 \hbar/J$ when working with $N\!\approx\!4$ internal states. 

Next, we remember that the $\Snd$ Chern number is extracted from the COM displacement along the $x$ direction [Fig.~2(b)]. It is thus relevant to estimate the typical displacement $x_{\text{c.m.}}(t^*)$ performed by the cloud during this critical time. Using Eq.~\eqref{velocities_appendix}, we find $x_{\text{c.m.}}(t^*)\!\approx\! a/2$ for $N\!\approx\!4$. Hence, in the absence of engineered periodic boundary conditions, the 2CN-measurement would necessarily require single-site resolution imaging; we note that a differential measurement would show a COM displacement of about one lattice site, $2x_{\text{c.m.}}(t^*)\!\approx\! a$. In order to overcome this technical difficulty, one should either use more internal states (we find $2x_{\text{c.m.}}(t^*)\!\approx\! 5a$ for $N\!\approx\!10$, which could be realized by populating the ground-state manifold of $^{40}$K atoms), or more preferably, engineer periodic boundary conditions~\cite{Celi:2014}, as already presented in the main text. \\

\section{Synthetic dimensions and anisotropy effects}

In this Section, we investigate the effects of anisotropy, which are inherent to Raman-induced synthetic dimensions for cold atoms~\cite{Celi:2014}. To address this question, we consider the relevant configuration of Ref.~\cite{Celi:2014}, in which Zeeman sublevels $\vert m_{F} \rangle$ in the hyperfine ground-state manifold (with total angular momentum $F$) are coupled by two Raman lasers, with frequency difference $\omega_1\!-\!\omega_2\!=\! \omega_{\text{R}}$ and wave-vector difference $\bs{k}_1 \!-\! \bs{k}_2\!=\!\bs{k}_{\text{R}}$. The energies of the Zeeman sublevels $\omega_{m_{F}}$ are shifted by a (real) constant magnetic field, and we assume that the resonance condition $\omega_{\text{R}}\!=\omega_{m_{F}+1}\!-\!\omega_{m_{F}}\!$ is fulfilled for all $m_F$ (i.e. we neglect detuning and quadratic Zeeman effects; see Ref.~\cite{Celi:2014}). This atom-light-coupling configuration is described by an effective Hamiltonian \cite{Goldman:2014bv}
\begin{align}
&\hat H_{\text{eff}}=\Omega_{\text{R}} \left (\hat F_{+} e^{i \bs{k}_{\text{R}} \cdot \bs{x}} +\hat F_{-} e^{-i \bs{k}_{\text{R}} \cdot \bs{x}} \right ) , \quad \hat F_{\pm} =\hat F_{x} \pm i \hat F_{y},  \notag
\end{align}
where we introduced the ladder operators acting on the ground-state manifold
\begin{align}
&\hat F_+ \vert m_F \rangle = g_{F,m_F} \vert m_F +1 \rangle, \\
&g_{F,m_F}=\sqrt{F(F+1) -m_F(m_F+1)},
\end{align}
and where $\Omega_{\text{R}}$ is the generalized (two-photon) Rabi frequency \cite{Goldman:2014bv}. This effective Hamiltonian can be interpreted as a tight-binding Hamiltonian describing tunneling along a synthetic dimension spanned by the $2F\!+\!1$ internal states,
\begin{align}
&\hat H_{\text{eff}}=\sum_{w=1}^{2F+1} J_w (w) e^{i \bs{k}_{\text{R}} \cdot \bs{x}} \, c_{w+1}^{\dagger} c_w + \text{h.c.} , \\
&J_w (w)=\Omega_{\text{R}} \sqrt{F(F+1) -m_w(m_w+1)},\label{eq:gs}
\end{align}
where $c_{w}^{\dagger}$ creates a particle at lattice site $w$, $J_w$ is the tunneling amplitude, and where $\exp \left (i \bs{k}_{\text{R}} \cdot \bs{x} \right)$ is the Peierls phase-factor exploited in the main text. In Eq. \eqref{eq:gs}, we introduced the coefficient $m_w\!=\!m_{F}\!$ corresponding to the internal state $\vert m_{F} \rangle$ living at site $w$; in the following, we use the convention according to which the first site $w\!=\!1$ hosts the lowest-energy state with $m_F\!=\!-F$, the next site $w\!=\!2$ hosts the state with $m_F\!=\!-F+1$, and so on. 

Atomic-lattice systems based on synthetic dimensions are potentially anisotropic \cite{Celi:2014}, due to the fact that the tunneling rates $J_w$ in Eq. \eqref{eq:gs} explicitly depend on the $w$ coordinate, through the coefficients $g_{F,m_F}$. A method by which anisotropy can be limited in the full lattice (including real and synthetic dimensions) consists in tuning the Rabi frequency to the value
\be
\Omega_{\text{R}}=J/ \langle g \rangle , \quad  \langle g \rangle =\sum_{m=1}^{2F}   g_{F,m_F}/2F, \notag
\ee
where $J$ is the hopping amplitude along the real dimensions (we assume that tunneling rates associated with real dimensions can be made homogeneous and isotropic). In this case, the tunneling amplitudes are given by 
\be
J_w(w)=J  \left ( \sqrt{F(F+1) -m_w(m_w+1)} / \langle g \rangle \right ) \approx J  \quad \forall w. \label{eq:defj}
\ee
We now investigate the tunneling rates using the configuration defined in Eq. \eqref{eq:defj}. For $F\!=\!1$, we have $g_{F,m_F}\!=\!\sqrt{2}$ for all Zeeman states, so that the system is genuinely isotropic in this case: $J_w \!=\!J$. In the case $F\!=\!3/2$, we obtain the following tunneling rates $J_w(w)$ between the four lattice sites:
\be
w=1 \xrightarrow{0.95 J} w=2 \xrightarrow{1.10 J} w=3 \xrightarrow{0.95 J} w=4 . \notag
\ee
In the case $F\!=\!9/2$ (ten sites), we find
\begin{align}
&w=1 \xrightarrow{0.71 J} w=2 \xrightarrow{0.95 J} w=3 \xrightarrow{1.09 J} w=4  \notag \\
&w=4 \xrightarrow{1.16 J} w=5  \xrightarrow{1.19 J} w=6 \xrightarrow{1.16 J} w=7 \notag \\
&w=7\xrightarrow{1.09 J}  w=8 \xrightarrow{0.95 J} w=9  \xrightarrow{0.71 J} w=10  \notag 
\end{align}

The effects of anisotropic tunneling rates $J_w$ was already studied in detail in Ref. \cite{Celi:2014}, for an atomic lattice made of one real and one synthetic dimension; this system exhibited the quantum Hall effect and was thus characterized by a band structure with nonzero Chern numbers. In that study, it was shown that the anisotropy due to $J_w$ only slightly modified the bulk bands, and in particular, that this imperfection preserved all topological gaps (even in the case of large $F$); see the Supplementary Material of Ref. \cite{Celi:2014}.  

We have performed a similar study for the 4D system considered in this work, and we verified that the topological band structure remains remarkably robust against anisotropy. In the ideal case of an isotropic 4D lattice [see Eq. (7) in the main text], the bulk energy spectrum depicted in Fig. 1(b) displays a non-degenerate lowest band. This lowest bulk band, which is associated with a non-zero 2nd Chern number, is separated from higher-energy bands by a gap of size $\Delta\!=\!1.32 J$ and is characterized by a reasonably small bandwidth  $W\!=\!0.43 J$. These values for the gap and bandwidth were obtained for an isotropic 4D system of size $40\!\times\!40\!\times\!40\!\times\!40$ with periodic boundary conditions. 
 In order to study the effects of anisotropy, we have computed the bulk energy spectrum for the anisotropic 4D system, including the realistic tunneling rates described by Eq. \eqref{eq:defj}, for the cases $F\!=\!3/2$ (four sites) and $F\!=\!9/2$ (10 sites); the corresponding system sizes are $40\!\times\!40\!\times\!40\!\times\!4$ and $40\!\times\!40\!\times\!40\!\times\!10$, respectively. Similarly to the 2D case discussed in Ref. \cite{Celi:2014}, we obtain reasonably small deviations from the ideal (isotropic) configuration. In the case $F\!=\!3/2$, we obtain a topological gap of size $\Delta\!=\!1.39 J$ and a bandwidth $W\!=\!0.36 J$. These values are to be compared with $\Delta\!=\!1.43 J$ and $W\!=\!0.31 J$, which correspond to the values obtained for an isotropic system with four lattice sites along $w$. This negligible modification of the topological band structure indicates that the simulations presented in the main text [Fig. 2], which were realized for an isotropic system with four lattice sites along $w$, satisfactorily capture the physics of a realistic (anisotropic) lattice configuration. For $F\!=\!9/2$, we obtain a gap $\Delta\!=\!0.90 J$ [and $W\!=\!0.84 J$], which is still reasonably large as far as the Chern-number measurement is concerned. From this analysis, we conclude that all topological properties described in the main text for the ideal isotropic lattice are well preserved in the realistic anisotropic lattice. We emphasize that the (2nd) Chern-number measurement entirely relies on the topological properties of the populated band and on the size of the bulk gap $\Delta$, which are both well preserved under anisotropy effects. 

Finally, we point out that periodic boundary conditions can be applied along the synthetic dimension, using an additional Raman transition \cite{Celi:2014}; this is relevant when dealing with a small number of internal states; see Section \ref{sect:open}. Importantly, the tunneling rate  associated with the closing link, i.e. $J_w(w=2F\!+\!1)$, can be controlled independently, which can be exploited to further reduce anisotropy. The results discussed above in this Section were obtained by setting $J_w(w=2F\!+\!1) \!=\!J$.